\documentclass[
reprint,
 amsmath,amssymb,aps,
]{revtex4-1}
\usepackage{graphicx,subfigure,color}
\usepackage{dcolumn}
\usepackage{bm}
\usepackage{hyperref}
\usepackage[mathlines]{lineno}
 
\newcommand{\be}{\begin{equation}}
\newcommand{\ee}{\end{equation}} 
\newcommand{\ba}{\begin{array}}
	\newcommand{\ea}{\end{array}}
\newcommand{\bea}{\begin{eqnarray}}
\newcommand{\eea}{\end{eqnarray}}

\newcommand{\half}{\frac{1}{2}}

\newcommand{\del}{\partial}

\newcommand{\vev}[1]{\langle #1 \rangle}

\renewcommand{\Im}{{ \rm Im}}

\begin{document}

\title{The  emergence of  Strange metal and Topological Liquid \\ near Quantum Critical Point in a  solvable model 
}
\author{Eunseok Oh, Taewon Yuk,  Sang-Jin Sin }
\email{sjsin@hanyang.ac.kr} 
 \affiliation{Department of Physics, Hanyang University, Seoul 04763, Korea.}  \date{\today}%
\begin{abstract}
 We discuss  quantum phase transition 
 by an exactly solvable model  in  the dual gravity setup. By considering the  effect of the scalar condensation on the fermion spectrum  near the  quantum critical point(QCP), we find that 
there is a topologically protected fermion zero mode  
associated with the metal to insulator transition.   
We  also show  that  the strange metal phase  with T-linear resistivity emerges  at  high enough temperature   as far as the gravity has a horizon.  
The phase boundaries  are calculated according to  the density of states,  giving   insights on structures of the phase diagram near the QCP. 
%
\end{abstract}
\maketitle

\paragraph*{\bf Introduction:}
 The quantum critical point (QCP) is believed to be a door to   understanding strongly correlated systems\cite{Sachdev:2011mz}.    There,  
the particle characters are lost due to the interaction but  the gravity dual description \cite{Maldacena:1997re,Witten:1998qj,Gubser:1998bc} may work due to the striking similarities between the QCP and  the black hole: both get the universality  by the apparent information loss, so that many different systems   look the same. Both can be assigned with the spectral functions and transport coefficients so that  it is likely that the we can identify the two  if they are the same\cite{Liu:2013yaa,Hartnoll:2016apf}.
However,   too  many informations are lost at the QCP so that information about the QCP itself  is not enough  
to identify  a physical system and the properties  {\it  off}  the QCP  is essential.    

Since the   deriving force in the phase transition near the QCP  typically is a symmetry breaking, we recently  studied  its effect on the fermion spectral function \cite{Oh:2020cym} numerically.  As a result, we   found   key  features of  quantum matters including  the Fermi arc,   flat band and nodal lines as well as the gap and pseudo gap. 
This is rather surprising since the topology is associated with clear band structure which usually become fuzzy  in the strongly interacting system. 
So it raises unavoidable  question: why
  this is related  to the phenomena of appearance of the Fermi liquid in some of most strongly correlated system like heavy fermions. 
Answering theses question would   provide  a new angle to the  study of the quantum matters with strong correlation.  

\begin{figure}[ht!]
	\centering
	\subfigure[ $g\Phi >0,$ Gap]
	{\includegraphics[width=2.8cm]{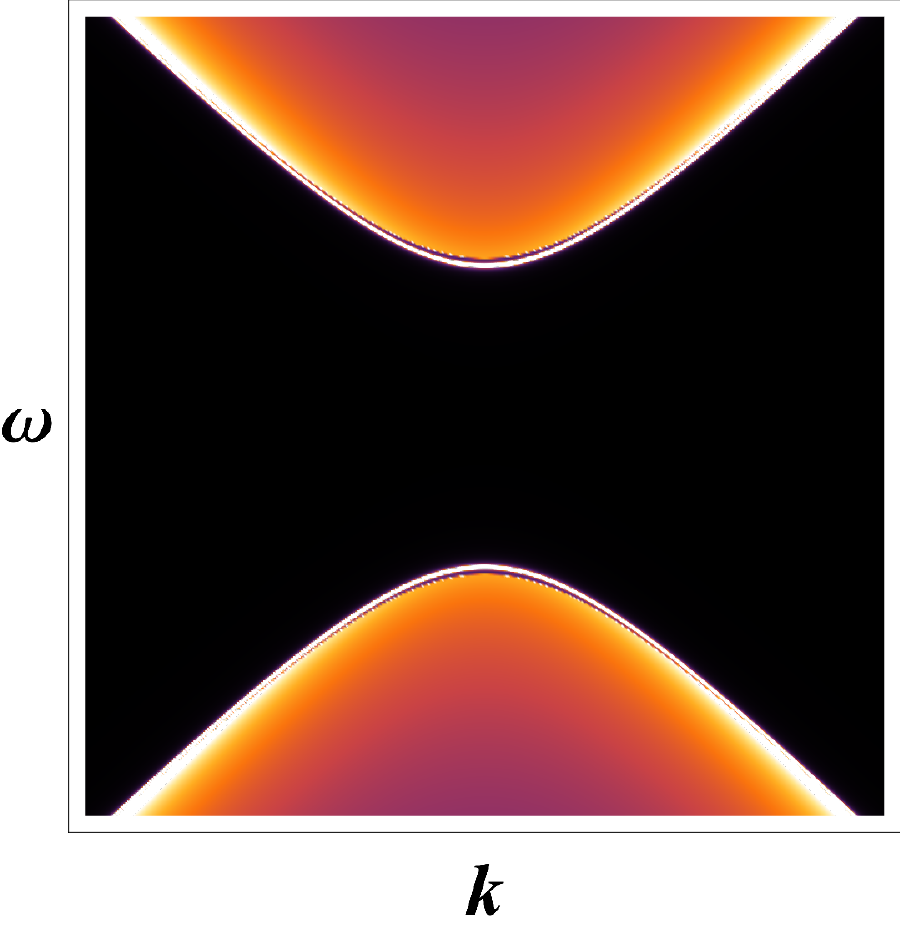}}
	\subfigure[ $g\Phi=0,$ QCP] 
	{\includegraphics[width=2.7cm]{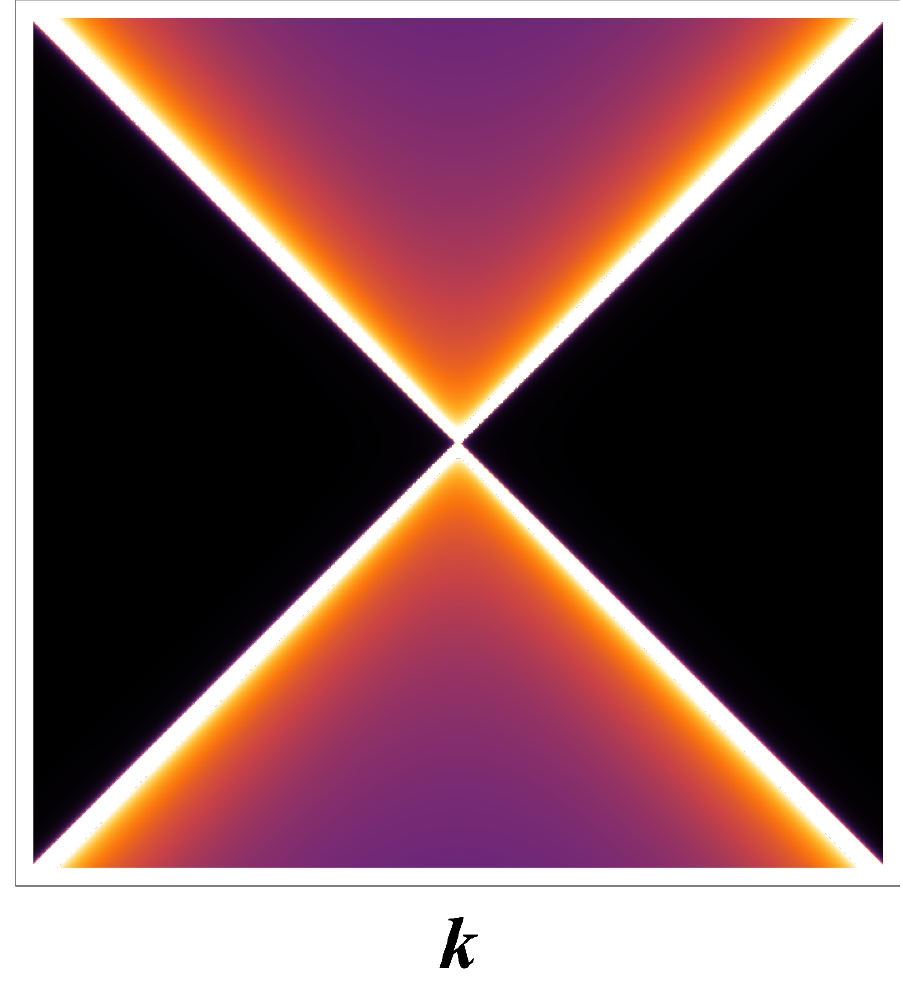}}	
	\subfigure[$g\Phi<0,$ Gapless]
	{\includegraphics[width=2.7cm]{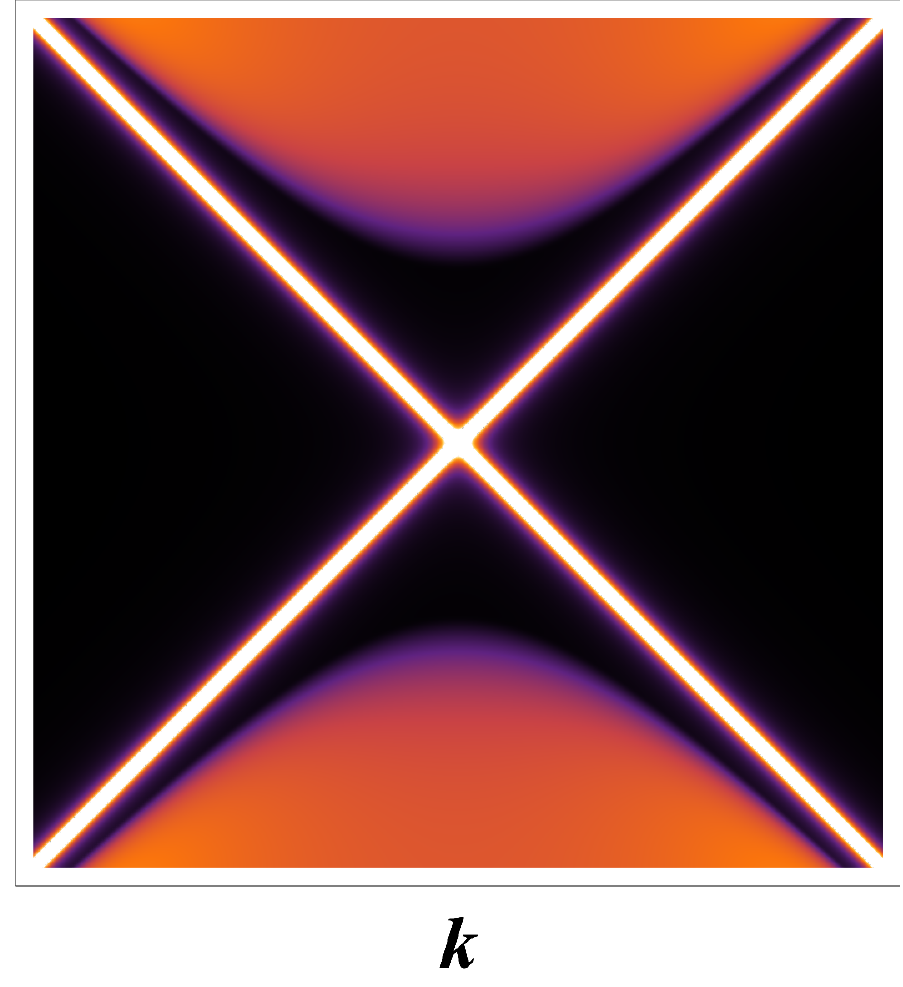}}
	\caption{Spectral Functions  (a) for  gapped phase, (b) for the  QCP,  (c) for the topological phase.  	}
	\label{scalar} 
\end{figure}

In this paper we  will show that   there can be    a gap or  a zero mode with topological stability depending on the sign of the scalar field. See Figure \ref{scalar}(a) and (c).  
In fact, the appearance of the gapless mode in spite of the chiral symmetry breaking order is a surprise because it can never happen in the flat space field theory. It is also curious question to ask if an ordered state is gapless,  how the order can be protected?    We will see that it is the topological property and its presence is related to the parity invariance  defined  in \cite{Oh:2020cym}.    

We will  first demonstrate  the presence of the   zero mode by solving the Dirac equation analytically. 
We   will then show  that the fermion zero mode has  the same mathematical structure with that of the topological insulator (TI).  It is the anti-de Sitter  space (AdS) version of the Jackiw-Rebbi solution\cite{Jackiw}. 
This also implies that  the gapless phase is a dissipation free, hence we   call  it  as the topological liquid (TL).  
Such  topological character of the zero mode makes  the TL different from the critical point in the spectrum:  in Figure \ref{scalar}(c) there is  clear separation between the gapped and    the zero mode, which is characteristically different from the critical point of Figure \ref{scalar}(b) which shows a gapless  fuzzy distribution of density of states.

The difference between the usual TI and and our system is that the zero mode in a usual TI describes a surface phenomena of the matter while it describes the bulk phenomena of the the physical matter here.

Because  both gap and gapless phases are created  out of a quantum critical point by a single order parameter with   different sign, we interpret that it   describes a phase transition at QCP from a (semi-) metal to insulator or its magnetic analogue. 
Indeed, when we turn on the temperature $T$  and 
calculate  the   spectral function as  a function of the $T$,  we  found   that the gapless phase has half width $\Gamma\sim T\sim 1/\tau$ which can be translated  as the  linear resistivity in $T$, signaling the presence of the strange metal.   
We will see that the so called Fermi liquid must exist in our theory as intermediate zone of the topological phase and the strange matter phase.
We also find that the gapped phase also  evolves to the strange metal as temperature goes high.
It turns out that the   presence of the zero modes is related to the  fact that the gravity dual use the asymptotically  anti de Sitter space which must  have a boundary, and  the appearance of the strange metal is associated with the presence of the black hole horizon.
 
An interesting aspect of our model is that the scalar  is associated with the chiral symmetry breaking in the holographic space which does not break any obvious   symmetry in the original space, which is a reminiscent of the  situation in the spin liquid. It can provide a possibility that 
some of the orders  which    traditionally has not been associated with  symmetry-breaking order may be associated with such order in higher dimensional holographic space. 
 Our theory is applicable to the class of materials with  transitions  from insulator to topological (semi-)metal \cite{PhysRevLett.109.186403,Salehi:2016aa,Yang:2014aa,Song:2019asj}. Finally, we  suggest that the stable particle provided by the zero mode can make the  Fermi liquid appear even  in  very strongly correlated system. 

\paragraph*{\bf The fermion zero mode with scalar in AdS:}
Let the bulk fermion $\psi$  be the  dual field  to the boundary fermion $\chi$ and $\Phi^{I}$ be  the  dual bulk field of the operator 
${\bar \chi}\Gamma^I\chi$.  We can  encode  the effect of the symmetry breaking on the spectrum of the fermions by consider the coupling  $\bar{ \psi}\,\Phi\cdot\Gamma\, \psi $ where $\Phi$ becomes a classical field under the symmetry breaking. 
The fermion equation of motion is given by 
 $
\left(  \Gamma^\mu{\mathcal{D}}_\mu   
-m-g\Phi \right)\psi =0
$
with $g=\pm 1$.   
The covariant derivative ${\cal D}_{\mu}$ is given by ${\cal D}_{\mu}=\partial_{\mu} +\frac14\omega_{\mu ab} \Gamma^{ab}$. 
In this paper we   use the simplest  AdS black hole metric whose explicit form is given in the supplementary material A. 
In the zero temperature,  $f=1$.   For $m_\Phi^2=-2$  the solution for the scalar 
is given by  
$ \Phi= M_{0}z +Mz^{2}$.   In Poincare coordinate at zero temperature, both $M_{0}$ and $M$ can be  independent boundary conditions, while for finite temperature with blackhole background, one of them is a boundary condition  and the other is function of the other.  Only when   $M_{0}=0$,   the value of $M$  is identified as the traditional condensation. 
The fact that $\Phi$ can include the non-zero source term  $M_{0}$ which correspond to the external driving force   is  the important  difference   of the 'order parameter field' in the holographic treatment.  Physically $M_{0}$ can be related to    the mass or doping parameter.   One can check that 
the zero temperature solution remains good approximation  even for finite temperature. This is  because   the  large $z$  region  where true solution deviates much from the probe solution is cut off  by the presence of the horizon.

We  can prove the existence of the   zero modes by solving above Dirac equation explicitly.  In   the supplementary material A, 
the readers can find the analytic form of the Green functions at zero temperature  whose pole give    spectrum. 
For $\Phi=M_{0}z$, 
by analyzing     the Green functions carefully we can see that  the   zero mode pole of $G^{g=1}_R$   is cancelled, while {\it   that of  $G^{g=-1}_R$   survives}.  The    massive particle spectra  is given by 
$
\omega^{2}-k^{2}=M_{0}^{2}\left(1- \frac{m^{2}}{(n+m+1)^{2}}\right) $  for  both  $g=\pm 1$ with  $n=0,1,2 \cdots$.  

Similarly, for $\Phi=Mz^{2} $, the spectrum is given by 
$
\omega^{2}-k^{2}=4M(n+m+1/2) $  { for } $g=1$ and 
$\omega^{2}-k^{2}= 4Mn$ { for } $g=-1 $. Notice that the first  spectrum is gapful for any $n$,  but  the  second one has a  zero mode at $n=0$.

 From these results, we see that   
for both $\Phi=M_{0}z$ and $\Phi=Mz^{2}$ cases,  two completely different phases, one gaped and the other gapless, can emerge   from the  same  quantum critical point at $g=0$ by invoking the scalar order depending the sign of the order parameter.
This is a surprising aspect of AdS space because such phenomena    never  happen in flat space or  in weakly interacting system where the scalar order introduces a gap, but not a zero mode.     We want to understand its origin and implications. 
 
\begin{figure}[t]
	\centering	  
	\vskip -.5cm
	\subfigure[Jackiw-Rebbi  mode]
	{\includegraphics[width=4.2cm]{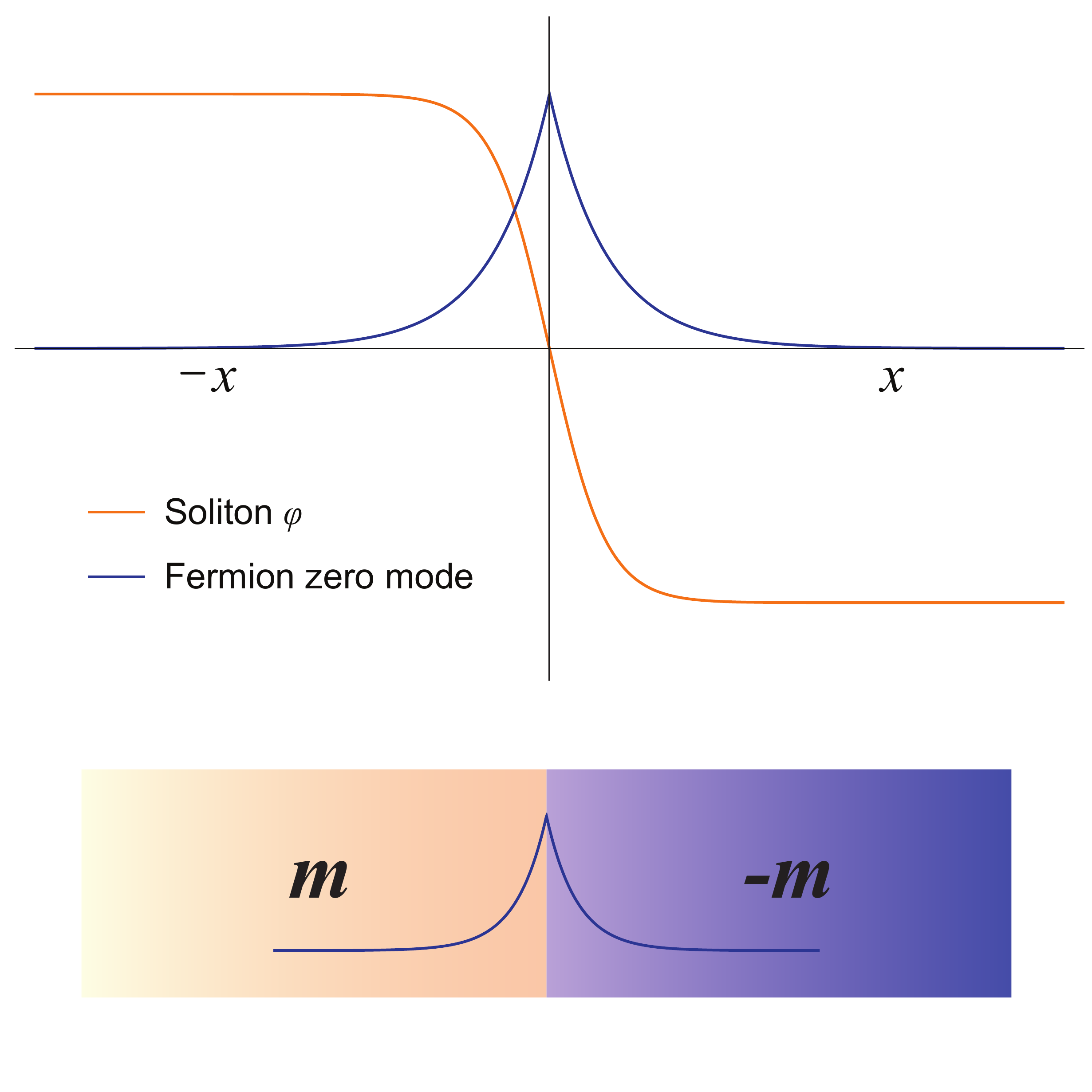}}
	\subfigure[JR mode in AdS]
	{\includegraphics[width=4.2cm]{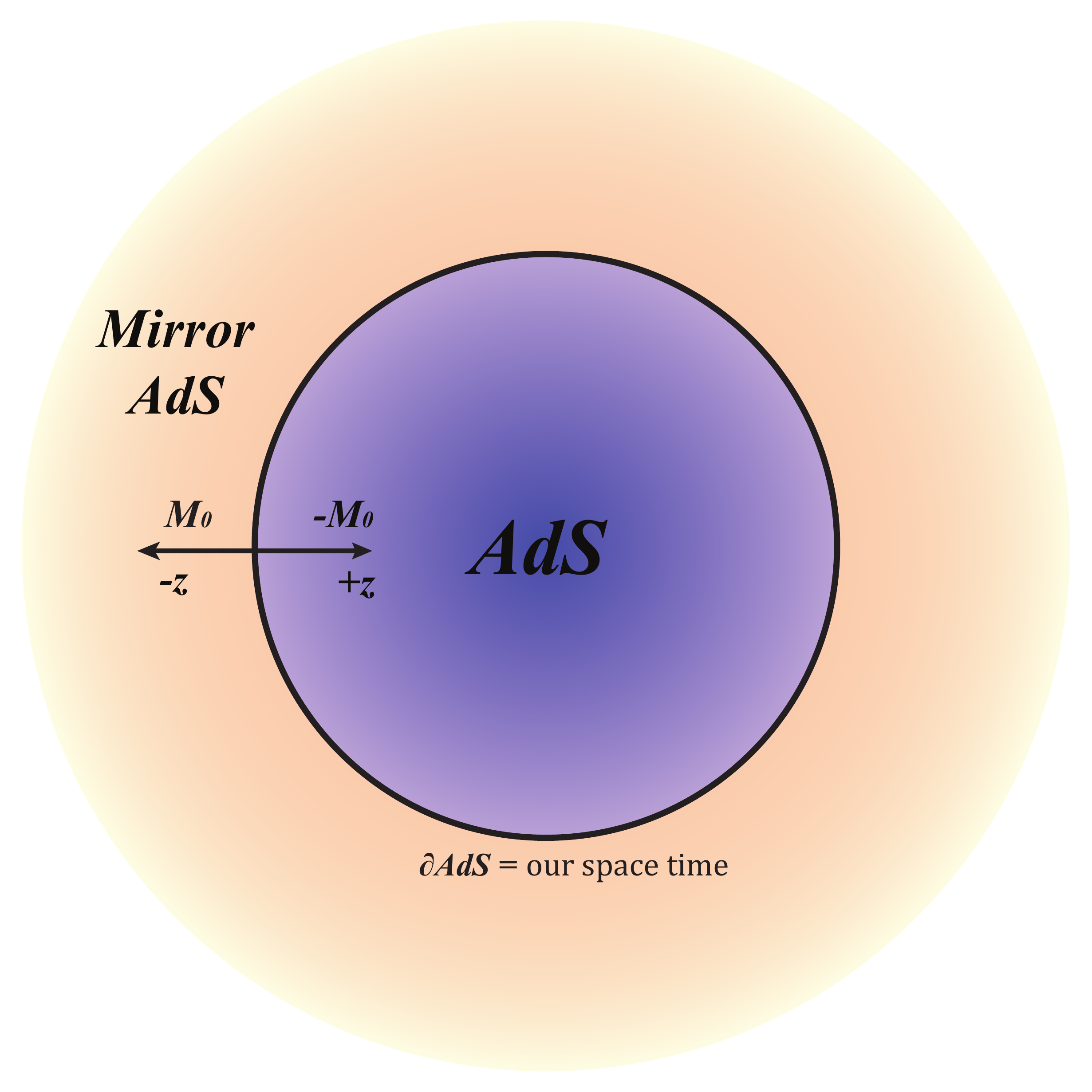}} 
	\caption{(a:top): Jackiw-Rebbi fermion zero mode in the soliton background. (a:bottom) Realization of the soliton $\varphi$ by  sign changing mass. 
		(b) $M_{0}$ corresponds to the $m$.
		The boundary of AdS  is the bulk of the physical space.  }
	\label{AdSTI}
\end{figure}

\paragraph*{\bf The topological insulator  in AdS:}
 The   closest phenomena is  the Jackiw-Rebbi (JR) fermion zero mode in the  soliton background \cite{Jackiw}. It is a solution of the Dirac equation 
$
(\gamma^{\mu}D_{\mu}-\varphi)\psi=0, 
$
where $\varphi$  changes sign across the domain wall. In such case  the fermion has a normalizable zero mode, $\psi_{0}(x)= \exp{(-\int dx \varphi )}$  localized at the domain wall. 
Its stability is guaranteed  by  the  boundary condition of $\varphi$ which makes $\varphi$  a topological soliton.  See Figure \ref{AdSTI}(a)top. 
Much   activities were performed under the name of the topological insulator after this solution is realized as the surface mode of condensed matter systems \cite{hasan2010colloquium,qi2011topological}. 
The  configuration of $\varphi$ can be realized by the  sign  changing    fermion mass   across   the boundary of the material,  $\varphi=m \cdot{\rm sign}(x)$. See Figure \ref{AdSTI}(a)bottom. 

Although we did not introduce any boundary of the physical system here,   the gravity dual description use asymptotically Anti-de Sitter (AdS) space which has a boundary, that is identified with the physical space where the material is sitting.  

Below, we will show that  our zero mode solution can be considered as the JR mode  in the AdS, which has a  boundary at $z=0$ and its bulk is  the region of $z>0$. 
The argument can be greatly simplified for  the pure AdS limit where the Dirac equation becomes 
\be
\left[ \Gamma^{z}\partial_{z}  -iK_{\mu}\Gamma^{\mu} +\frac{m+g\Phi}{z}  \right] \phi =0, 
\label{eqMs1}
\ee 
$\hbox{ with } K_{\mu}=(-\omega,k_{x},k_{y}).$
Now we consider    $\chi_{0 }(k)$ with  $k=(\omega,{\bf k})$ satisfying $ K_{\mu}\Gamma^{\mu} \chi_{0}(k)=0$,  
which is  nothing but the Dirac equation with zero mass at the boundary of the AdS. 
Then  $\chi_{0\pm}(k)$ defined by $ \Gamma^{z}\chi_{0\pm}(k)=\pm \chi_{0\pm}(k)$ are also solutions. 
The zero mode in AdS can be constructed from this by   
$\phi_{0}(z,k)= {\tilde\phi}(z) \chi_{0}(k)$, where $\tilde\phi$ is a scalar satisfying 
 the eq.(\ref{eqMs1}) for $K_{\mu}=0$. 
The solution  is  given by 
$
\phi_{0}(z,x)=   z^{-m} \exp(-g\int_{0}^{z} dz' \varphi(z') )\chi_{0+}+
z^{m} \exp(g\int_{0}^{z} dz' \varphi(z') )\chi_{0-}, \label{zeromode}
$
where $\varphi=\Phi/z$ for $z>0$. 
The  standard (alternative) quantization choose $\chi_{0+}$ ($\chi_{0-}$) \cite{Iqbal:2009fd,laia2011holographic}.  
By choosing $m<0$ and $g=-1$,   the wave function in the bulk are normalizable and localized at $z=0$. 
One should  also notice that   $z^{\pm m}$ factor does not  make an issue for the normalizability because 
$|m|<1/2$ by the unitarity bound\cite{Iqbal:2009fd}. 

To make the parallelism with  Jackiw-Rebbi solution describe above, we    introduce the mirror AdS in the regime  $z<0$. See Figure \ref{AdSTI}(b).  
The sufficient condition for the normalizability of the zero mode in $-\infty<z<\infty$,  is 
$\varphi(-z)=-\varphi(z)$ which is clearly satisfied by  
 \be
\varphi=M_{0}~{\rm  sign}(z)+Mz.
  \ee
It is instructive to consider the effect of each term in  $\varphi$.    
Corresponding fermion zero modes are 
$\psi_{-0}^{(M_{0})}= |z|^{m} \exp(- M_{0}|z|)  \chi_{0-}$ 
and $\psi_{-0}^{(M)}= |z|^{m} \exp(- \half Mz^{2})  \chi_{0-}$ respectively. They are  zero modes localized at the boundary for $m<0$. 
The reader can read  more explicit  proof   in the supplementary material E. 
 Since the boundary of the AdS is the physical space bulk, our zero mode is the bulk mode of the real material  unlike the TI in the weakly interacting system. 
 
 Summarizing, when there is a scalar condensation in the strongly interacting system, our theory predicts that there should be a topological liquid with non-dissipative zero mode, giving the metal or semi-metal depending on the size of the Fermi surface,  which in turn  can be tuned by the chemical potential.  


%

\paragraph*{\bf Phase diagrams near  the QCP:}
Since  both the gap and the gapless features are created  out of a  QCP by a single order parameter,  there is a metal insulator transition at the QCP. 
Then, by adding the temperature,   
we  can  discuss   the phase diagram near the quantum critical point. 
The dual of finite temperature is described by the   black hole geometry.   But we can  use order parameter field which is the solution for for pure AdS  as a leading approximation.  
To see the typical density of states(DOS) of each phases, we calculated the DOS numerically.  Interested readers can see the   Figure S1 of the supplementary material.

The density of state depends on the order parameter and  temperature, 
and the shapes are quantitatively parametrized by the half width's dependence on the two parameters. 
Therefore we can classify the phases according to  $\Gamma$ as function of $T/M_{0}$.  
We can get the analytic result for it so that 
the entire phase diagram can be understood analytically.   

\begin{figure}[ht!]
	\centering	  
	\subfigure[$\Gamma(T)$ at $g M_0=-2$]
	{\includegraphics[width=4cm]{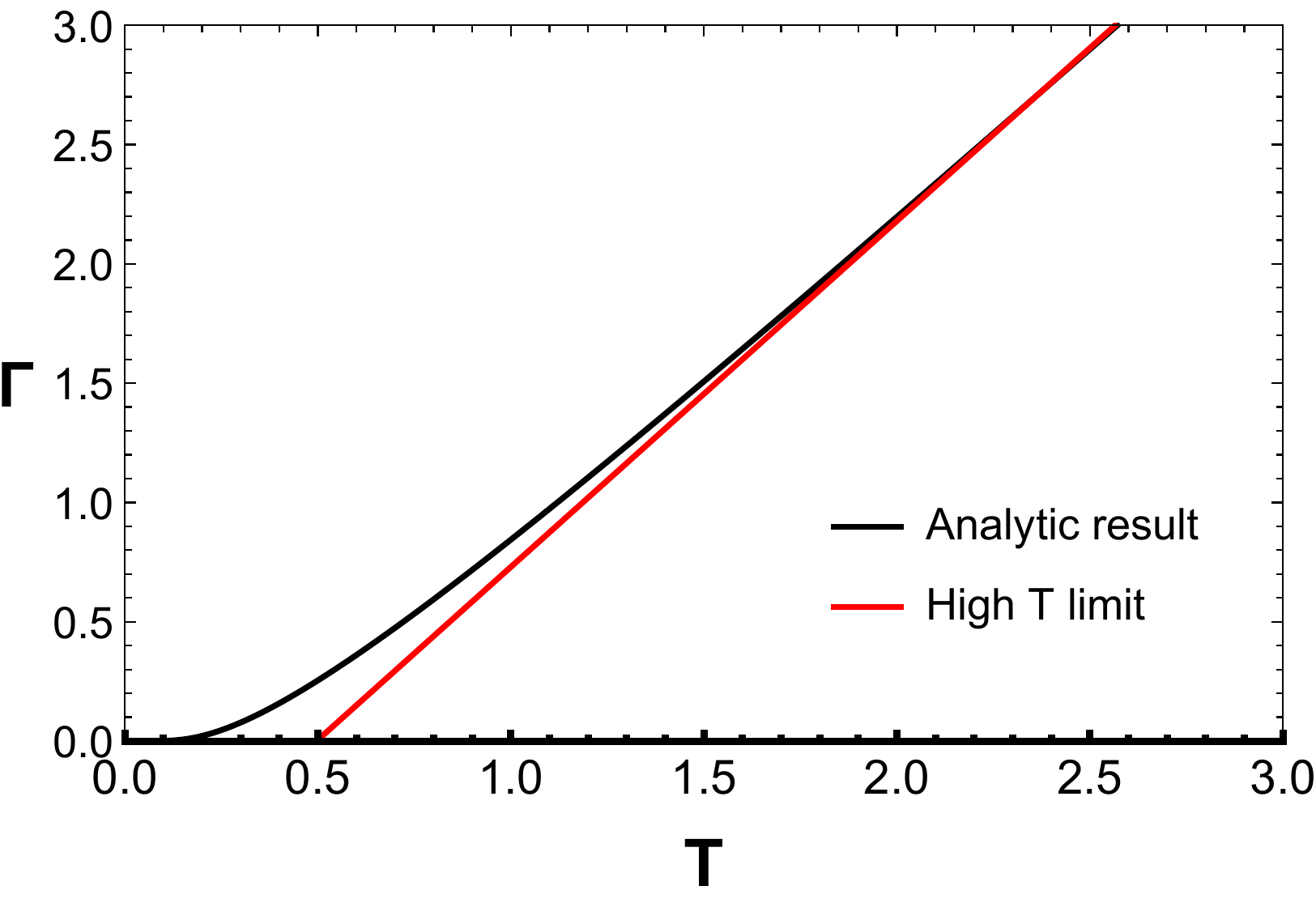}}
	\subfigure[$\Gamma(T)$ at $g M_0=+1$]
	{\includegraphics[width=4cm]{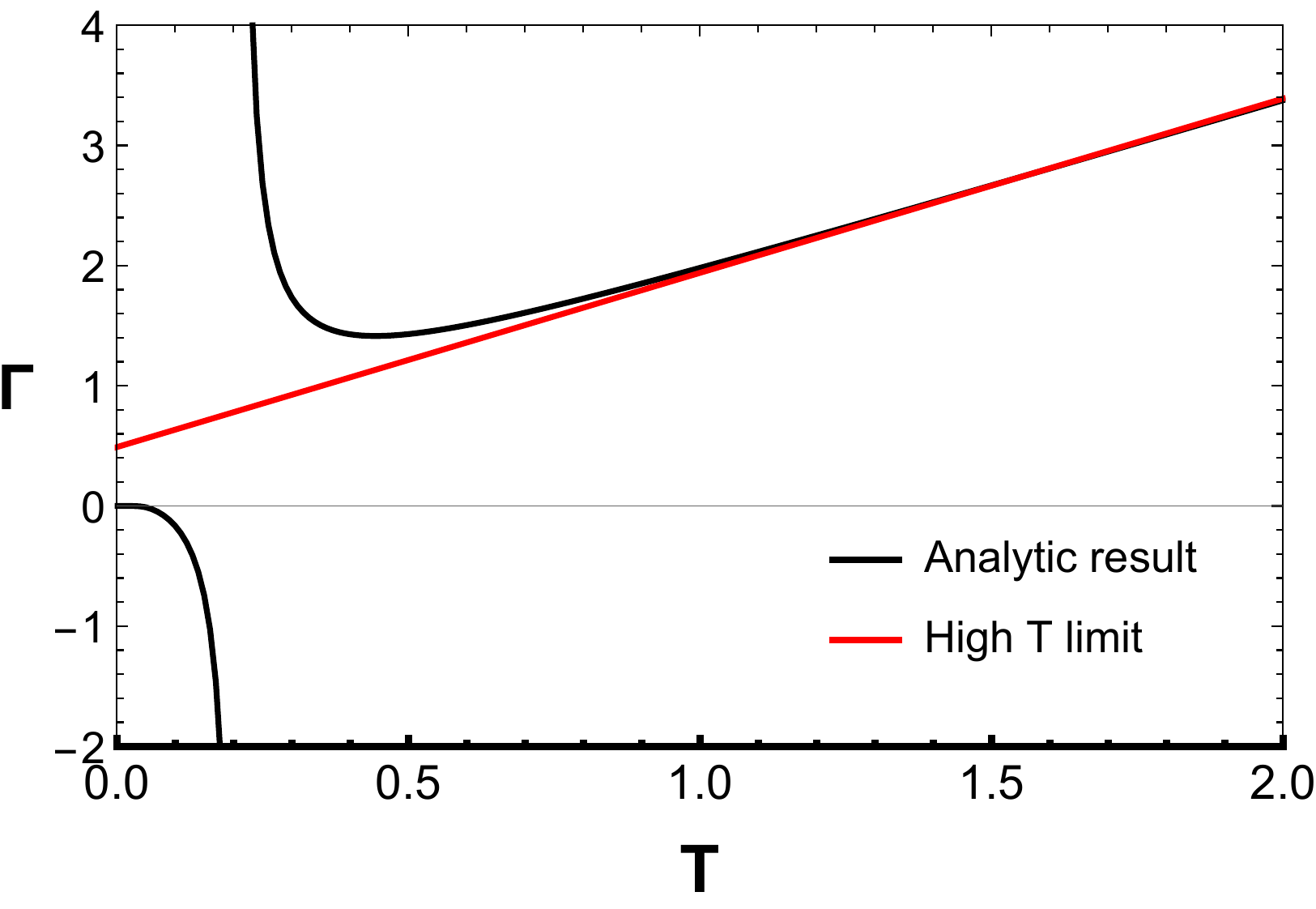}}
	\caption{ (a,b) $\Gamma$ as a function of $T$. 
		For high temperature, $\Gamma(T)\sim T $  universally.   In (b) negative $\Gamma$ at low $T$ means the   appearance of the gap there.}
	\label{spectralshape}
\end{figure}

{\it The half width $\Gamma(T)$} of fermion density of state   for  AdS$_{4}$  with $\Phi=M_0 z +M z^2$ is given by, 
\be
\Gamma (T)= {2 \pi T}/{ \int_{0}^{1}  \frac{dt F(t)}{t (1-t)^{2/3} }},    \label{gammaT3}
\ee
with {\small $
	F(t)=-\sinh [ \frac{4m}{3}\tan^{-1}\!\!\sqrt{t}+ {g M_0 \beta_{0}(t)}/{T}+  {g M \beta_{1}(t)}/{T^2 }] $}
where $\beta_{0}(t)=\frac{1}{2\pi}B(t;\frac{1}{2},\frac{1}{3})$ and $\beta_{1}(t)=\frac{3}{8 \pi^{2}}B(t;\frac{1}{2},\frac{2}{3})$. Here $m$ is the bulk mass of the  Dirac fermion in AdS$_{4}$, and in this paper we take $-1/2<m<0$.  The derivation and more general result can be found in the supplementary materials C. 
Notice that when the temperature is much larger than the order parameter, $
\Gamma\simeq\pi T/\gamma_{m}, $ 
with 
$\gamma_{m}^{-1}=\frac{2 m}{3} B(\frac{1}{2},\frac{1-2m}{3}) {}_3F_{2}(\frac{1}{2},\frac{1}{2}-\frac{2m}{3},1-\frac{2m}{3};\frac{3}{2},\frac{1}{2}+\frac{1-2m}{3};1)$.  We will give  general argument later that this implies the appearance of the strange metal phase later.  
In the Figure  \ref{spectralshape}, we plotted the half width in eq.(\ref{gammaT3}) as function of $T$ for $gM_{0}=-2$. 

\begin{figure}[ht!]
	\centering	  
	\subfigure[structure of PD for  $M_{0}$]
	{\includegraphics[width=4cm]{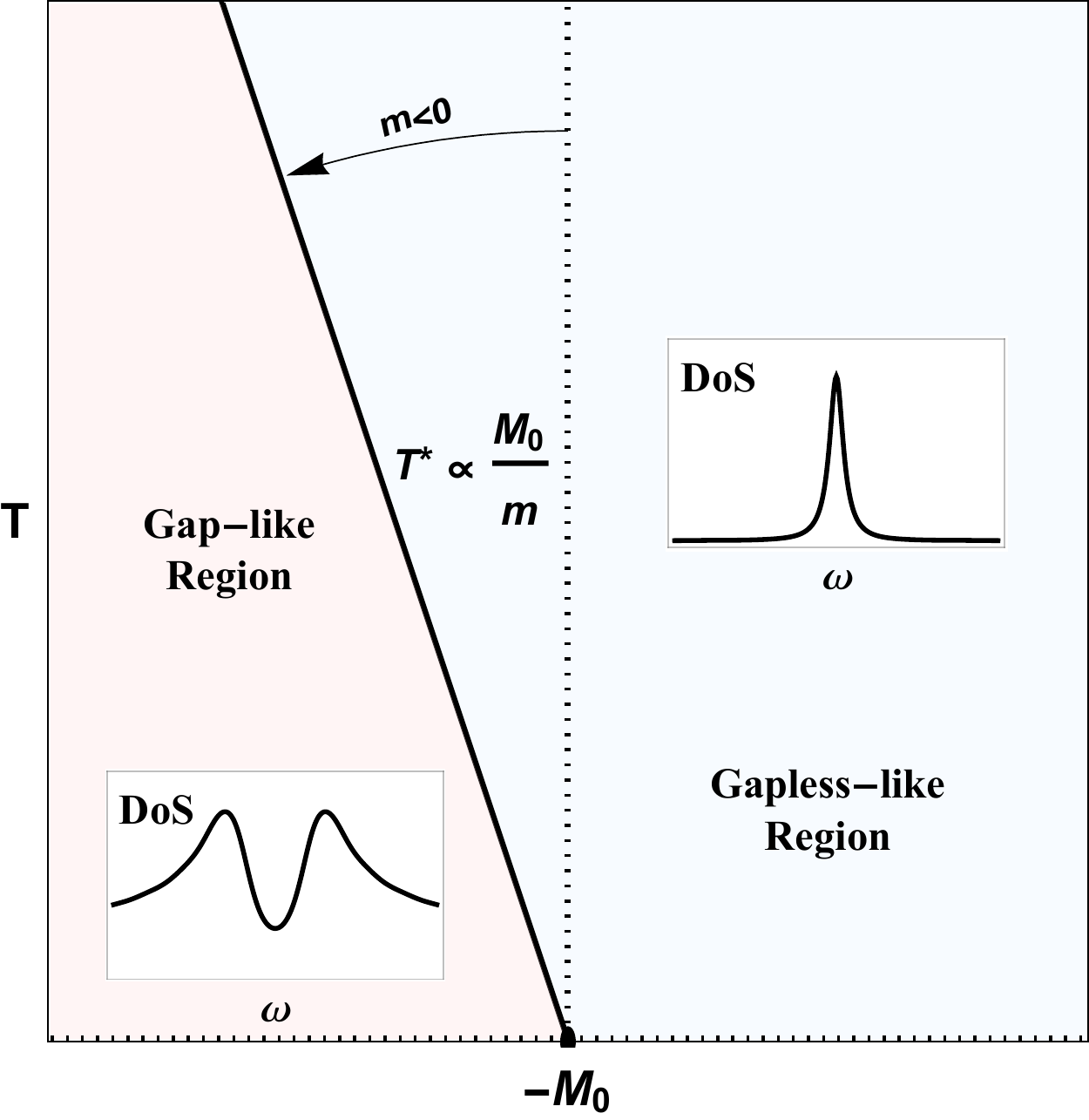}}
		\hskip 0.4cm
	\subfigure[phase diagram with $M_{0}$]
	{\includegraphics[width=4cm]{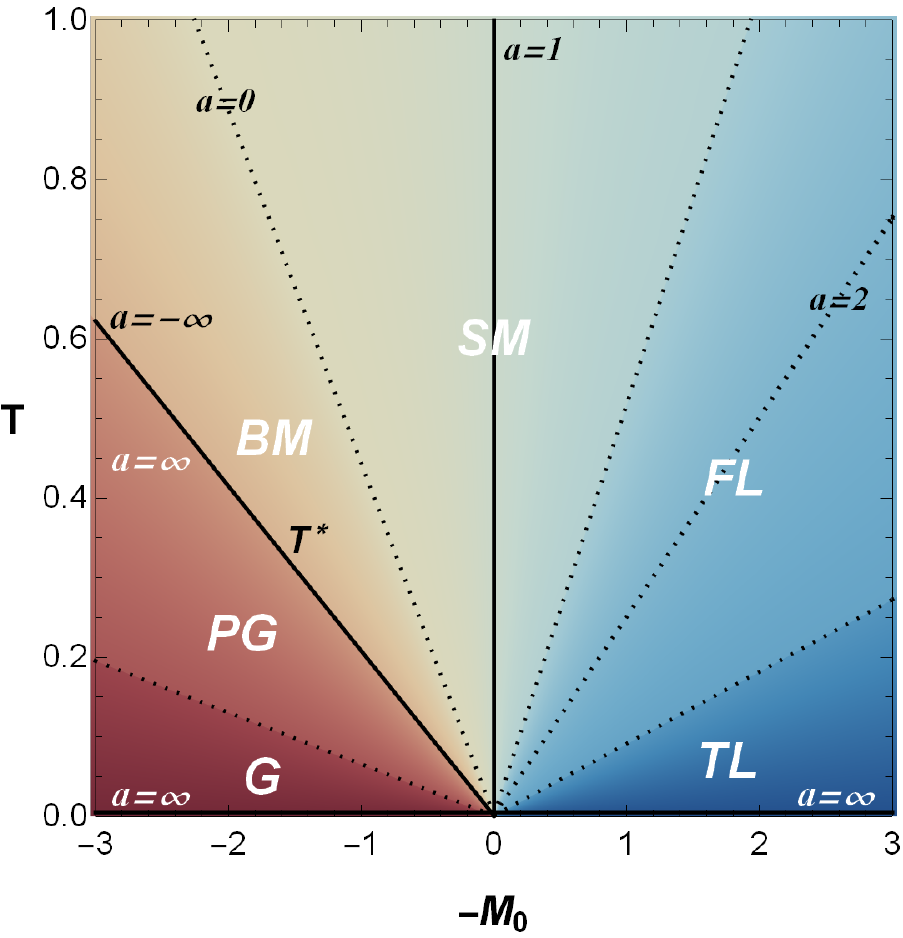}}	  
\\
	\subfigure[structure of PD for  $M$]
	{\includegraphics[width=4cm]{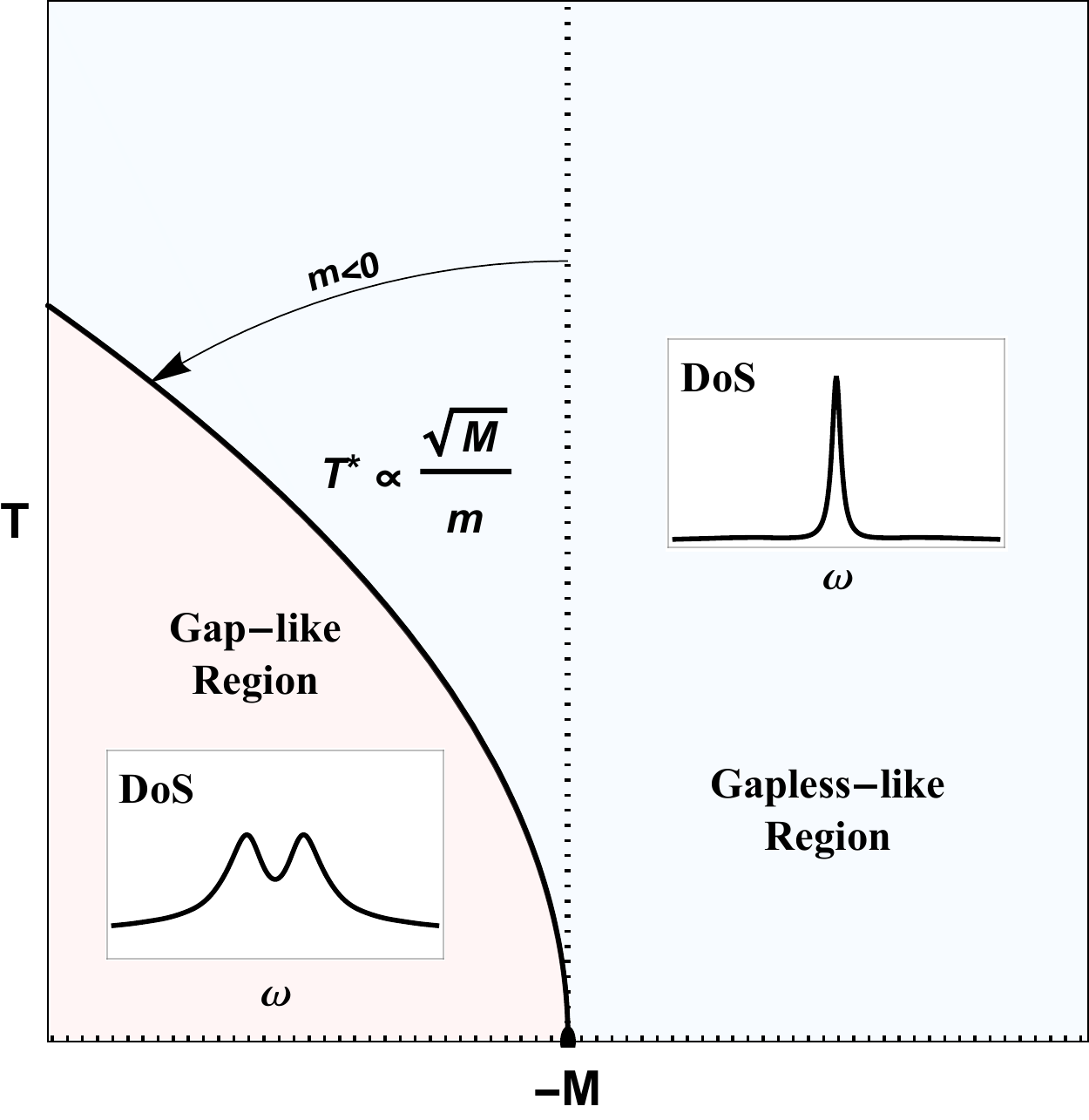}}
	\hskip.4cm
	\subfigure[phase diagram with $M$]
	{\includegraphics[width=4cm]{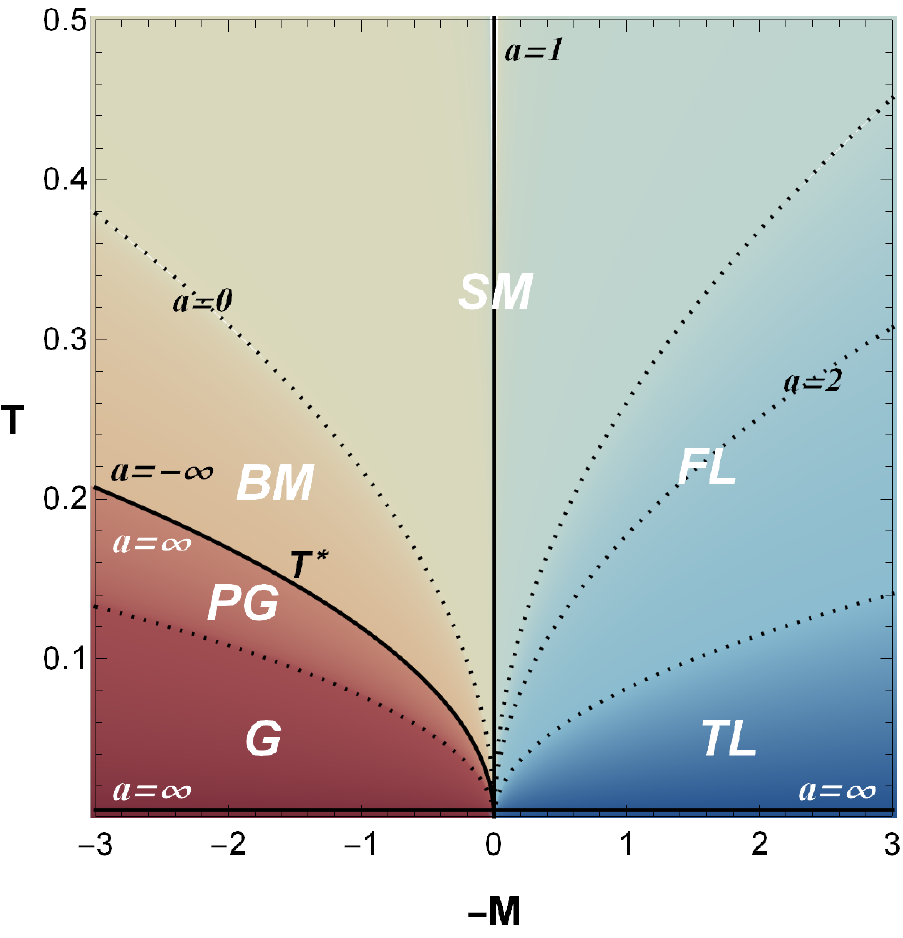}}	  
	\caption{  	
	True Phase transition  exists only along  $T_*$ lines.  Others are crossover. 
	 In 	(b,d)    `Fermi liquid(FL), strange metal(SM) and bad metal(BM)' phases are around $a=2,1,0$ lines respectively.   }
	\label{msteer1}
\end{figure}

{\it  Phase boundaries}  can also be calculated by using  eq. (\ref{gammaT3}). 
The phase diagram in   $(M_{0},T)$ plane is consequence of the competition of  three regimes:  
i) Topological liquid  whose core is   the negative $M_{0}$ axis at $T=0$,  ii)Gapped  insulating phase  whose core is  at the positive $M_{0}$ axis. 
iii) the strange metallic phase whose center is defined by $a(T/M_{0}):=\frac{\del \log \Gamma}{\del \log T}=1$. It is along $T$ axis. We emphasize that these three lines are the center of the phases  not  phase boundaries. 
There is one true phase boundary in this phase diagram and it is  $T^{*}$ line given by $T^{*}=M_{0}/m$ with $-1/2<m<0$, 
which is boundary between gapped and gapless regimes. 
The Figure \ref{msteer1}(a) explains this idea.   
Notice that $M_{0}$ is not an order parameter but a source parameter so that it can be a parameter that can describe a phase diagram. See the Figure \ref{msteer1}(b).

Now the important point is that for $\Phi=M_{0}z$, $a(T,M_{0})=$constant lines where $\Gamma(T)\sim T^{a}$   are   straight lines as it can be   seen from eq.(\ref{gammaT3}). As we rotate it starting from positive  $-M_{0}$ axis counterclockwise, $a$ move from $\infty$ to $-\infty$ arriving at $T^{*}$ line. 
It should first pass the $a=2$ line where $\Gamma\sim T^{2}$, hence the  core of the Fermi liquid phase.   Then it pass the $a=1$ where  $\Gamma\sim T$,  the core of the strange metal phase.   
Upon the crossing the $T^{*}$ line,    $a$  jumps from  $-\infty$ to  $\infty$. 
As we rotate further it decreases 
to  arrive at the minimum  and then increases to arrive at the negative $-M_{0}$ axis which is the core of insulating phase. 

We define the phase boundary of the gapped and pseudo gap phases by the line at which $a$ is minimum. 
The resulting phase diagram is Figure \ref{msteer1}(b). 
 If we assume  that  the  strength of the source $M_{0}$ is related with the doping rate $x$ by $M_{0}\propto (x_{0}-x)$,  then our phase diagram mimic the that of the typical phase diagram near the QCP. 
 
So far we confined ourselves to the order associated with explicitly broken  symmetry   with 
$\Phi=M_{0}z$ with $M=0$.  
We can repeat the analysis for the  
$\Phi=Mz^{2}$ with $M_{0}=0$.   What is important is to 
notice that $\Gamma/T $ depends on the $M$ only through the combination  $M/T^{2} $, and so is 
the exponent function $a(T)=\frac{d \log \Gamma}{d \log T}$. 
This means that   all the phase boundary lines follow 
$
T\propto \sqrt{M}$. We remark that we should  consider $M$ as a boundary condition, not as a condensation. The latter is determined by setting $M_{0}=0$.   The result is the Figure \ref{msteer1}(c,d).


%

\vskip .3cm
\paragraph*{\bf Emergence of the strange metal  and the Fermi liquid with strong correlation}
One can understand the emergence of the strange metallicity 
$\Gamma(T)\sim T$ directly from the equation of the motion (\ref{eqMs1}).  By rescaling $z=\zeta z_{H}$, the equation  
contains the temperature   only in $\omega/T$ and  inside 
$\Phi=(M_{0}z_{H})\zeta+(Mz_{H}^{2})\zeta^{2}$. 
For   simplicity we consider the case $\Phi=M_{0}z_{H} \zeta $ only. 
When $M_{0}/T<<1$    the presence of the order can be neglected. Then the  $T$ dependence   comes only through $w:=\omega/T$. If the   spectral width in $w$  in this case  
is $\Delta w=\Gamma_{0}$, then $\Gamma(T)$ defined as the spectral width in $\omega$ is given by 
$
\Gamma(T)=\Delta \omega =T\Delta w=  \Gamma_{0} T. 
$
Therefore $\Gamma(T)$ is linear in $T$ whatever is $\Gamma_{0}$.
This explains the appearance of the strange metal  for high temperature region in Figure \ref{spectralshape}(c):    $\Gamma(T)\sim \hbar/\tau\sim T$  
can be translated  into the resistivity data $\rho\sim 1/\tau \sim T $.  

Notice that the same argument  can be applied     to the  equation describing the gauge field   fluctuation  in AdS space, and it gives  the linear temperature dependence for the width of the Drude-like peak  in the AC conductivity, which are  also linear in $ T$. 
It works as far as there is a horizon in the background gravity. This is the origin of the emergence    of the strange metallicity. 

When there is no order parameter, the strange metal appears even for $T\to 0$ limit, which explains why the strange metallic phase is of  fan shape starting from the QCP in the phase diagram.  
The strange metal's transport behavior has been discussed in the context of the holographic theory previously \cite{Faulkner:2009wj,Faulkner:2010da,Blake_2013,Davison_2014,Blake_2015,Ge:2016lyn}.  Our point is that apart from the existence of the  black hole horizon  and the temperature dominance over the order,  we do not need anything else to prove it.  

For low temperature regime, the kinetic term   is  very small  compared with the interaction term:  $| iK_{\mu}\Gamma^{\mu}  |<< M_{0}\zeta$, therefore $M_{0}$ become  the dominating  scale  and 
the system behavior follows the zero temperature physics. 
The zero mode that is responsible to  our gapless mode is the   Jackiw-Rebbi zero mode  which has topological stability  as we discussed before.  
This predicts the emergence of the topological liquid near zero temperature.  
The topological nature   causes    unusual stability of the Fermi-surface and it  is expected that there is no dissipation in this phase.    Such stability of the zero mode as a particle spectrum also predict that there should be a Fermi liquid phase in the system. One should  also remember  that 
the   Fermi liquid phase  in our theory appears 
as an intermediate region between the topological liquid and the strange metal. In heavy fermion system   Fermi liquid is observed in spite of the strong correlation  making the effective mass of the fermion   1000 times heavier. It would be interesting if we  can connect our observation to   such   extraordinary stability of the Fermi liquid in terms of real data.  We hope to come back to this issue soon.  
  
  \begin{acknowledgements}
This  work is supported by Mid-career Researcher Program through the National Research Foundation of Korea grant No. NRF-2021R1A2B5B02002603 and by the BK21 FOUR Project in 2020. We  
thank the APCTP for the hospitality during the focus program, “Quantum Matter and Quantum Information with Holography”, where part of this work was discussed.
\end{acknowledgements}

 \bibliography{Refs_scalar}

\onecolumngrid
\newpage

\section*{Supplementary materials for}


\begin{center} 
 \vskip 1cm
{\Large \bf  The  emergence of  Strange metal and Topological Liquid in a solvable model of   Quantum Phase Transition:}
 
  \vskip 1cm
 {\large 
  {Eunseok Oh,} 
 {Taewon Yuk, }
  {Sang-Jin Sin}$^{*}$
}
 \vskip 0.2cm
 { Department of Physics, Hanyang University, Seoul 04763, Korea }
  \vskip 1cm

 \end{center}
 \appendix 
 
 \renewcommand\thefigure{S\arabic{figure}}    
 \setcounter{figure}{0}   
  \setcounter{equation}{0}   
 \renewcommand{\theequation}{S\arabic{equation}} 

\section{Spectrum and the zero modes  with scalar order for $g=-1$}
Here we show the presence of the zero mode by working out the full spectrum. 
Our fermion action is given by the sum $S=S_{g,A,\Phi}+S_{ \psi}+S_{bdry} $,   where  	
 \bea
S_{g,A,\Phi}&=&  \int d^{d+1}x\sqrt{-g} \left( R+\frac{6}{L^2}-\frac{1}{4}F^2_{\mu\nu} +D_{\mu}\Phi_{I}^2 -m^{2}_{\Phi}\Phi^{2}\right),   \\
		 S_{ \psi} &=&\int d^{d+1}\sqrt{-g} x\;  i\bar{ \psi}
		\left(\half \Gamma^\mu(\overrightarrow{\mathcal{D}}_\mu  -\overleftarrow{\mathcal{D}}_\mu )
-(m+g\Phi) \right)\psi, \, \\
S_{bdy}&=&\pm\frac{i}{2}\int_{\partial M}d^dx\sqrt{-h}\bar{\psi}\psi.
 \label{action} 
 \eea
This action give the complete dynamics of all the fields including 
the Dirac field. 
		According to the choice of sign of $S_{bdy}$, half of the bulk spinor degrees of freedom are projected out so that  for $+(-)$ sign,   only 	
	$\psi_{+}(\psi_{-})$ survive and  this choice of the boundary action is called  standard (alternative) quantization\cite{Faulkner:2013bna,laia2011holographic}. 		
In this paper we   use the simplest  AdS black hole metric,    
\bea
	d s^2=-\frac{1}{L^2}\frac{f(z)}{z^2}d t^2+\frac{L^2}{z^2f(z)}d z^2+\frac{1}{L^2 z^2}d x_{i}^2,\quad   
	f(r)=1-(\frac{{z}}{z_H})^{d-1},
\eea
where the horizon radius  is related to the temperature by $z_H= d/4\pi T$ and we set $L=1$. 
If we define $\phi_\pm(z)$ by 
 $\psi_{\pm}=(-\det g g^{zz})^{-1/4}e^{-i w t + i k_i x^i}\phi_\pm(z)$,  
 the  $\phi$ satisfies 
\be
\left[ \partial_{z} +\frac{1}{\sqrt{f}}\Big(iK_{\mu}\Gamma^{\mu} +\frac{m+g\Phi}{z}   \Big)\Gamma^{z}\right] \phi =0, 
\hbox{ with } K_{\mu}=(-\omega/\sqrt{f},k_{x},k_{y}).
\label{eqMs}
\ee 
 Following the standard dictionary of AdS/CFT for the $p$-form bulk field $\Phi$ dual to the operator $O$ with dimension $\Delta$, its mass is related to the operator  dimension by 
\be 
m^{2}_{\Phi}=-(\Delta-p)(d-\Delta-p),
\ee
and asymptotic form near the boundary is 
\be 
\Phi=\Phi_{0}z^{d-\Delta-p}+ \vev{O_{\Delta}}z^{\Delta-p} .
\ee   
We use   following Gamma matrices representation \cite{Liu:2009dm},   
	\be
	 \Gamma^{r}=\sigma^{3}\otimes \mathbf{1}_{2}, \quad \Gamma^{\mu}=\tau^{1}\otimes \gamma^{\mu},   \hbox{ with } \gamma^{\mu}=\{i\sigma^{2}, \sigma^{1}, \sigma^{3} \}, \hbox{ for  }\mu=0,1,2 .
	\ee

 \subsection*{Case 1:  $\Phi=Mz^{2} $ } 
First we consider the case    $\Phi=Mz^{2}$.  Then the Dirac equation is equivalent to 
 \bea \label{FEM}
-\phi_{\pm}&''+\left(M^2 z^2+\frac{m(m \pm 1)}{z^2}+2gM(m\mp\frac{1}{2})\right) \phi_{\pm}={\cal E} \phi_{\pm},
  \eea
 where $ {\cal E}= {w^2-\vec{k}^2}$.
The solution to the eq.(\ref{FEM}) is given by 
\bea 
\phi_+&=2^{\frac{1}{2}(\frac{1}{2}- m)}z^{- m}e^{- M\frac{z^2}{2}}\left(\mathbf{C}_{1+} U^{\frac{1}{2} - m}_{u}( M z^2)+\mathbf{C}_{2+} L^{-\frac{1}{2}- m}_{-u}(M z^2)\right)\\
\phi_-&=2^{\frac{1}{2}(\frac{1}{2}+ m)}z^{m}e^{- M\frac{z^2}{2}}\left(\mathbf{C}_{1-} U^{\frac{1}{2}+m}_{v}(M z^2)+\mathbf{C}_{2-} L^{-\frac{1}{2}+m}_{-v}(M z^2)\right)\\
\hbox{with }\varepsilon &=\frac{w^2-\vec{k}^2}{4 M},  u=\frac{1}{2}(g-1)(m-\frac{1}{2})-\varepsilon,  \; v=\frac{1}{2}(g+1)(m+\frac{1}{2})-\varepsilon , \\
U_{u}^{k}(z)=&z^{-u}{}_{2}F_{0}(u,1+u-k;;-z^{-1}), \; L_{u}^{k}(z)=\frac{\Gamma(k+1+u)}{\Gamma(k+1)\Gamma(u+1)}{}_{1}F_{1}(-u,k+1;z)
\eea
where $\mathbf{C}_{i\pm}$ are two component constant spinors and  $U_{u}^{k}$ and   $L_{u}^{k}$ are   associated Laguerre, whose asymptotic behavior
 determines the  normalizability of $\psi$.
Since the Laguerre function  $L_{u}^{k}$ in general  contains 
$e^{M z^2}$ we need to set $\mathbf{C}_{2\pm} =0$.  
The $z\rightarrow \infty$ behaviors are 
\bea
\nonumber
	\phi_{+}&\sim{2^{\frac{1}{4}-\frac{m}{2}} e^{-\frac{M z^2}{2}} z^{-2 u-m}  \mathbf{C}_{1+}M^{-u} }, \\
	\phi_{-}&\sim{2^{\frac{1}{4}+\frac{m}{2}} e^{-\frac{M z^2}{2}} z^{-2 v+m}  \mathbf{C}_{1-}M^{-v} }.
\eea

Then,  $z\rightarrow 0$ behaviors are given by  
\bea
\phi_+ \sim 2^{\frac{1}{4}-\frac{m}{2}}\left(z^{-m}B_{1+}+z^{m+1}B_{2+}\right), \quad
\phi_- \sim 2^{\frac{1}{4}+\frac{m}{2}}\left(z^{1-m}B_{1-}+z^{m}B_{2-}\right)
\eea
{\small 
\bea
 \hbox{ where  }\quad 
B_{1+}&=\mathbf{C}_{1+} \frac{\Gamma(m+ {1}/{2})}{\Gamma(u+m+ {1}/{2})},   \quad \quad\quad\quad\;\,
B_{2+} =\mathbf{C}_{1+}M^{ {1}/{2}-m}\frac{\Gamma (- {1}/{2}-m ) }{\Gamma(u)},
\cr
B_{1-}&=\mathbf{C}_{1-}M^{ {1}/{2}-m}\frac{\Gamma(m- {1}/{2})}{
	\Gamma(u+m+ {g}/{2})}, \quad 
B_{2-}=\mathbf{C}_{1-}\frac{\Gamma\left( {1}/{2}-m\right) }{\Gamma(u+ {(g+1)}/{2})}.
\eea
}
The relation 
\bea
B_{2-}=i (2m+1) \frac{\gamma^\mu k_\mu}{w^2-{k}^2} B_{2+},\quad B_{1-}=i (2m-1) \frac{\gamma^\mu k_\mu}{w^2-{k}^2} B_{1+}
\eea
which was established in \cite{Liu:2009dm} still hold here in the presence of the interaction term.
Then the Green Function $G_R$ is defined by 
\bea
G_R=-i S \gamma^0, \quad \hbox{ with S defined by }
B_{2-}=S B_{1+}. 
\eea
Now we  can write down     Green functions  for each sign of $g$.
\bea
G^{g=1}_R&=&M^{-1/2+m}\frac{  \Gamma(\frac{1}{2}-m)\Gamma(\frac{1}{2}+m-\varepsilon)}{2\Gamma(\frac{1}{2}+m)\Gamma(1-\varepsilon)} \gamma^\mu k_\mu \gamma^t , \\
G^{g=-1}_R&=&M^{-1/2+m}\frac{ \Gamma(\frac{1}{2}-m)\Gamma(-\varepsilon)}{2\Gamma(\frac{1}{2}+m)\Gamma(\frac{1}{2}-m-\varepsilon)}\gamma^\mu k_\mu \gamma^t .
\eea
The poles of the Green function are given by those of gamma function at the  non positive integers  so that the spectra are given by  
\bea
\omega^{2}-k^{2}&=&4M(n+m+1/2),  \hbox{ for } g=1, \\
\omega^{2}-k^{2}&=&4Mn, \hbox{ for } g=-1 ,  \label{KKM}
\eea
with  $n=0,1,2, \cdots$. The first  spectrum is gapful for any $n$,  but  the  second one has the zero mode at $n=0$.

  \subsection*{ Case 2: $\Phi=M_{0}z$} 
Now we turn to the case     $\Phi=M_0z$. 
The equation of motion for $\phi$ with scalar source $M_0$ is equivalent to  
\bea  \label{FM0}
-\phi_{\pm}&''+\left(\frac{m(m \pm 1)}{z^2}+M_0^2 +g\frac{2  m M_0}{z})\right) \phi_{\pm}={\cal E}\phi_{\pm}.
 \eea
After fixing the coefficients to remove the divergent pieces in  $z \rightarrow \infty$ limit, the solution is 
\bea
\phi_\pm&=& e^{-\sqrt{\mu}z}(2\sqrt{\mu}z)^{\mp m}U_{\mp m + g\nu}^{\mp 2m}(2 \sqrt{\mu}z)\\
\mu&=&k^2-w^2+M_0^2, \quad 
\nu=\frac{ m }{\sqrt{1-\varepsilon'}}, \quad \varepsilon'=\frac{\omega^{2}-k^{2}}{M_{0}^{2}}.
\label{FM1}
\eea
 The $z\rightarrow0$ behavior of the solution is
\bea
\phi_{+} &\sim z^{-m} \frac{ (2\sqrt{\mu})^{-m}\Gamma(1+2m)}{\Gamma(1+m+g \nu)}+z^{1+m}  \frac{(2\sqrt{\mu})^{1+m}\Gamma(-1-2m)}{\Gamma(-m+g \nu)}\\
\phi_{-} & \sim z^{1-m} \frac{(2\sqrt{\mu})^{1-m}\Gamma(2m-1)}{\Gamma(m+g \nu)}+z^{m} \frac{(2\sqrt{\mu})^{m}\Gamma(1-2m)}{\Gamma(1-m+g \nu)}.
\eea
These data give the Green functions 
 for  $\Phi=M_{0}z$: 
\bea
G^{g=1}_R &=&  \frac{  (4{\mu})^{\frac{1}{2}+m} 
\Gamma (-2 m)\Gamma \left(1+m+\nu\right)}{\left(k^2-w^2\right)
	\Gamma(-m+ \nu )\Gamma (1+2 m)},
 \\
G^{g=-1}_R &=& \frac{   (4{\mu})^{\frac{1}{2}+m} \Gamma (-2 m)\Gamma \left(1+m-  \nu\right)}{\left(k^2-w^2\right)
	 \Gamma \left(-m-  \nu\right)\Gamma (1+2 m)} .
	 \label{GRFM0}
\eea
where parameters $\mu,\nu$ are given by 
 $\mu=k^2-w^2+M_0^2$ and   
$\nu=\frac{ m }{\sqrt{1-\varepsilon'}}$ with   $ \varepsilon'=\frac{\omega^{2}-k^{2}}{M_{0}^{2}}$.
Although these two look similar, there is a striking difference: notice that $ (k^2-w^2 )
	 \Gamma(-m+ \nu ) \approx -2M_{0}^{2}/m$ near the lightcone $k^{2}=\omega^{2}$, therefore the apparent zero mode pole of $G^{g=1}_R$   is cancelled, while {\it the zero mode  of  $G^{g=-1}_R$   survives}.  The    massive particle spectra exist only for $m<0$ if $g=1$, while  they    exist only for $m>0$ if $g=-1$. 
In both cases,   the massive tower is given by 
\be
\omega^{2}-k^{2}=M_{0}^{2}\left(1- \frac{m^{2}}{(n+m+1)^{2}}\right), \;\; n=0,1,2 \cdots. \label{KKM1}
\ee	
The spectra given in eq. (\ref{KKM}) and  (\ref{KKM1}) are the Kaluza Klein tower associated with the box character of  AdS space, which gives an effective compactification. 
  
  \section{ Typical  Density of States for various phases }
\begin{figure}[ht!]
	\centering	  
	\subfigure[Topological Liquid]
	{\includegraphics[width=3.7cm]{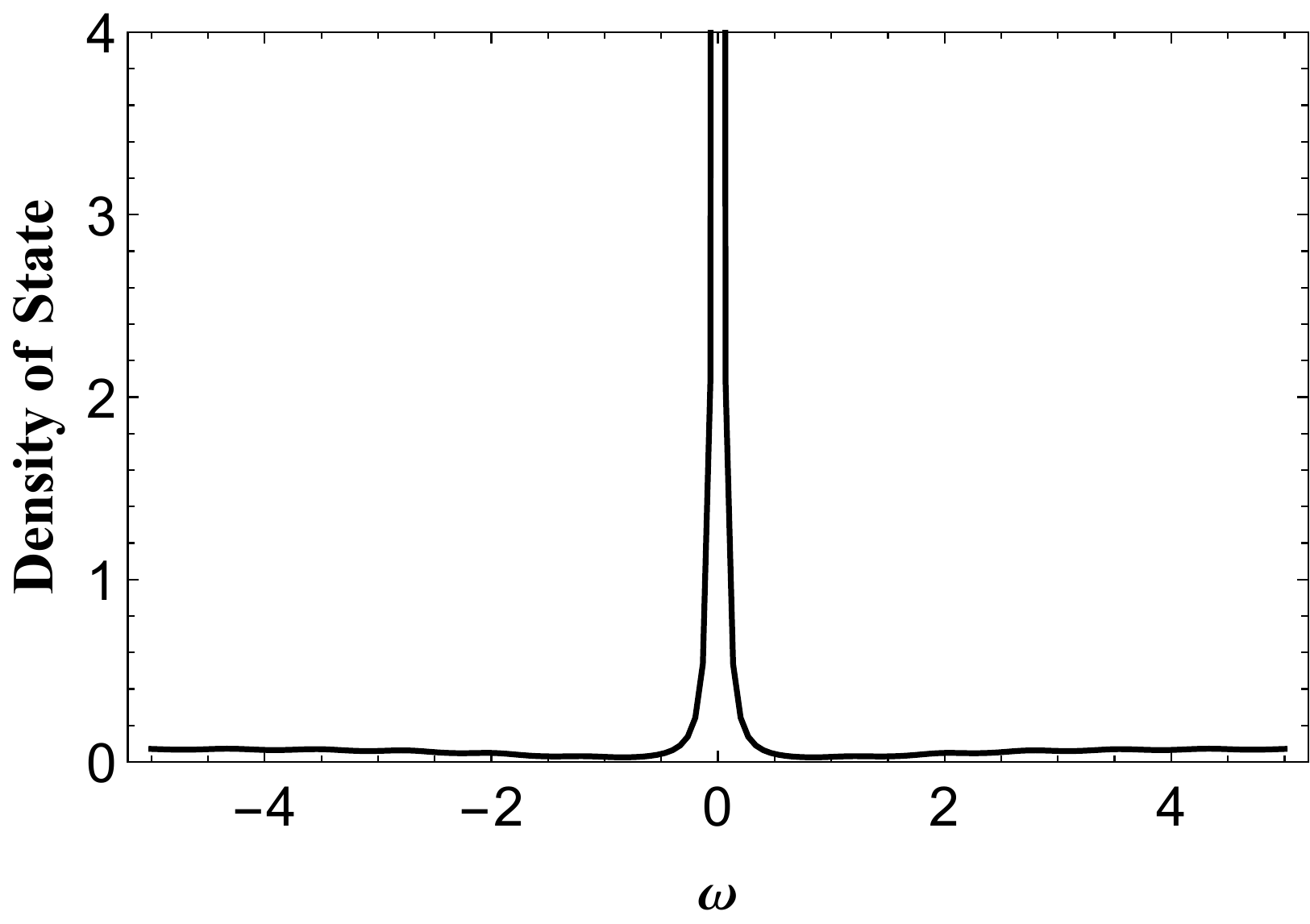}}
	\subfigure[Strange Metal]
	{\includegraphics[width=3.7cm]{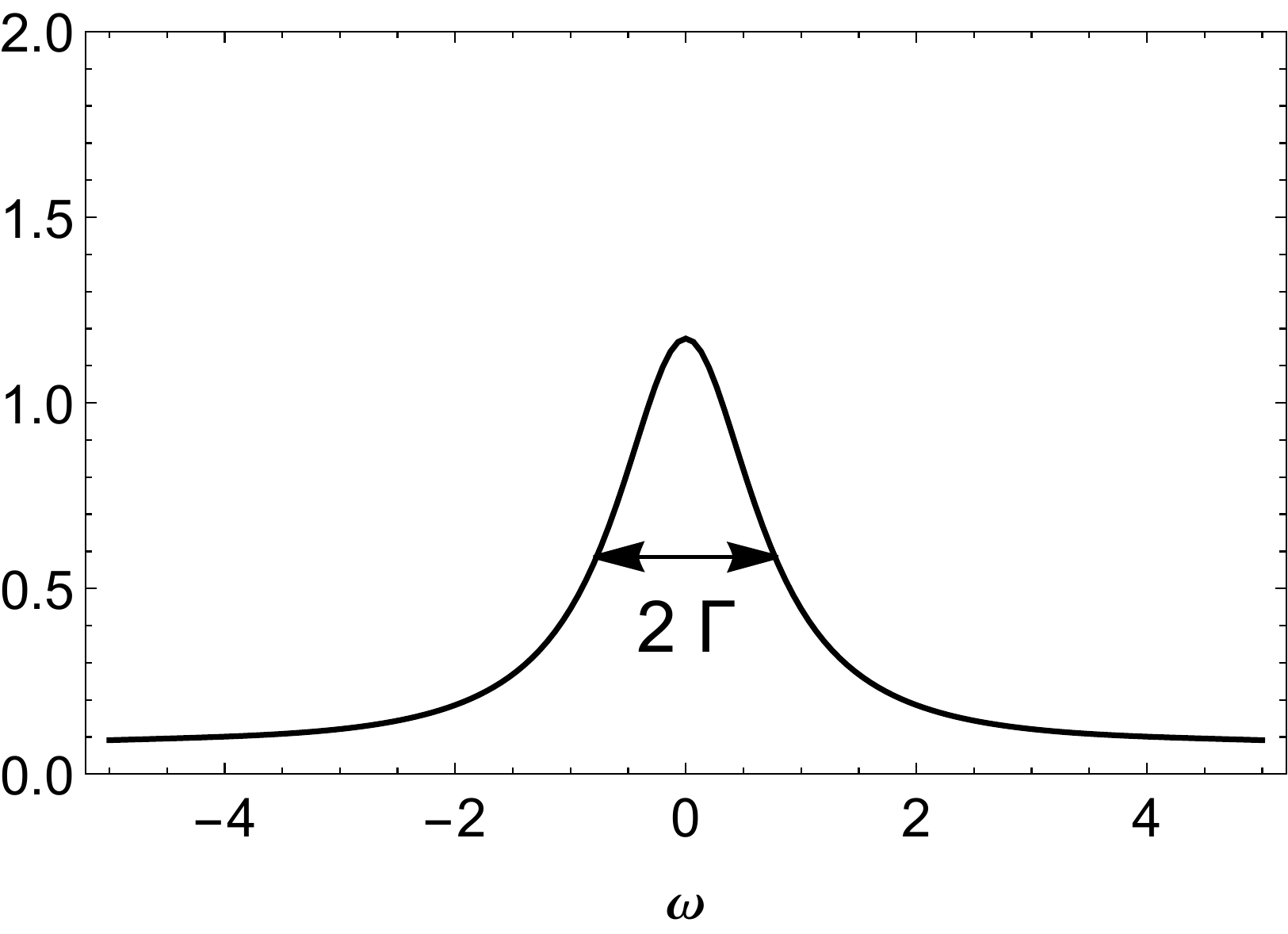}}
	\subfigure[Gap]
	{\includegraphics[width=3.7cm]{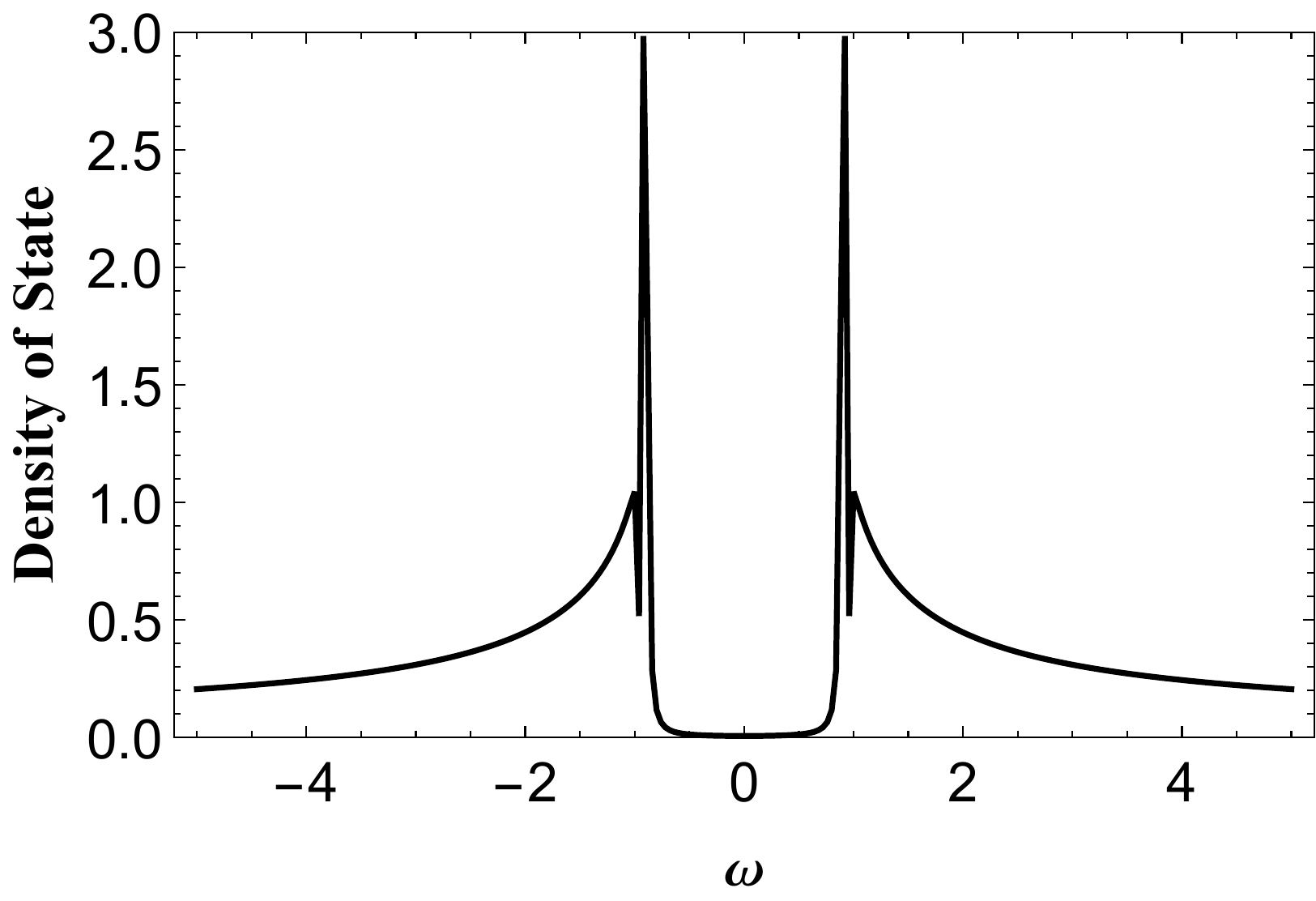}}
	\subfigure[Pseudo-gap]
	{\includegraphics[width=3.7cm]{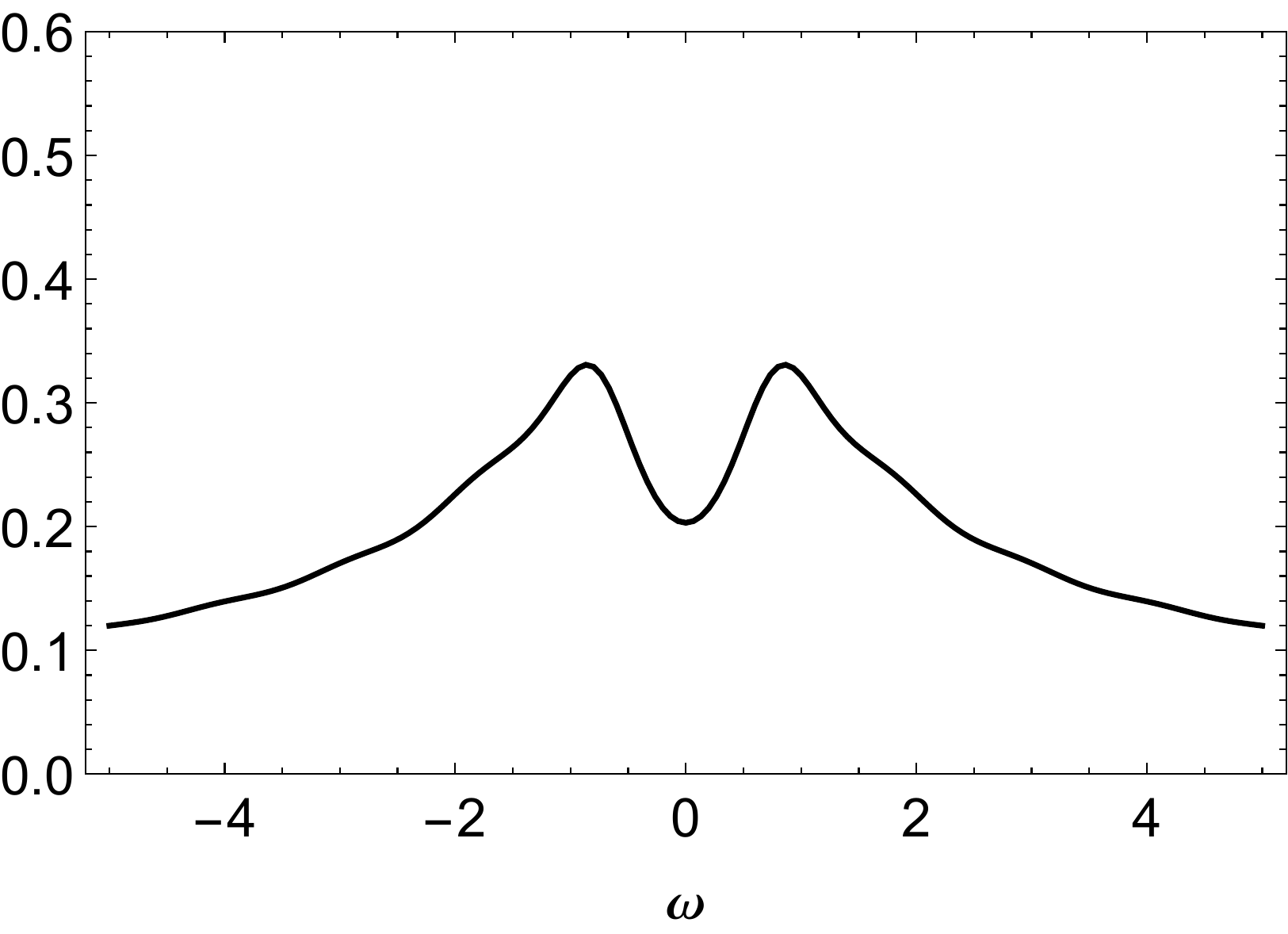}}
	\caption{DOS for different phases $\Gamma(T)$ for $m=-0.3$:  For (a,b)   $T=0.1, 0.8$ respectively along $g M_0=-2$. 
		For (c,d)  $T=0.01, 0.15$ respectively along $g M_0=1$.  In Figure (c) $\Gamma<0$     for  the gap.}
	\label{spectralshape1}
\end{figure}

  To see the typical density of states(DOS) of each phases, we calculated the DOS numerically.
 The Figure \ref{spectralshape1} is result of this. 

  \section{  Analytic   expression for  $\Gamma(T)$}
To calculate the $\Gamma$ as a function of the temperature, 
  we   define the spectral function $\cal A$ by 
\bea
\mathcal{A}=\textrm{Tr}(\textrm{Im} G_{R})
\eea
and   expand $\mathcal{A}$ for small $w$ by,
\bea
 \mathcal{A} &=  \mathcal{A}^{(0)}(1- \tau w + \cdots)\\
 &= \mathcal{A}^{(0)}+w \mathcal{A}^{(1)}+\cdots
\eea
Then the relaxation time $\tau$ is defined by the small $w$ expansion 
\be
\tau = \frac{1}{\Gamma}= -\frac{\mathcal{A}^{(1)}}{\mathcal{A}^{(0)}}
\ee
The Dirac equation \ref{eqMs} is equivalent to the flow equation for $\xi_{\pm}$ as follows
\bea
\label{floweq}
\left(\partial_z + 2\frac{g\Phi + m/z}{ \sqrt{f(z)}}\right)\xi_{\pm}(z)=\left(\frac{w}{f(z)}\pm\frac{k}{\sqrt{f(z)}}\right)\xi_{\pm}(z)^2+\left(\frac{w}{f(z)}\mp\frac{k}{\sqrt{f(z)}}\right)
\eea
where $\xi_+ =\frac{i y_-}{z_+}, \xi_- =-\frac{i z_-}{y_+} $ and $\phi_{\pm}= (y_\pm,z_\pm)^T$. 
The retarded Green function $G_R= z^{2m} {\rm diag}(\xi_+,\xi_-)|_{z \to 0}$\cite{Liu:2009dm}. From now on we set $k=0$, so that the equations for $\xi_+$ and $\xi_-$ are the same for $k=0$ and we delete the lower index $\pm$ from $\xi_{\pm}$.  Now, we expand the $\xi$ in $w$ up to first order,
\bea
\label{exp}
\xi(z) = \xi^{(0)}(z)+w \xi^{(1)}(z)+\cdots .
\eea
Substituting the eq. \ref{exp} into eq. \ref{floweq}, the equations for the zero-th and first order in w expansion becomes
\bea
&(\partial_z + 2\frac{g\Phi + m/z}{ \sqrt{f(z)}})\xi^{(0)}(z)=0 , \\
&(\partial_z + 2\frac{g\Phi + m/z}{ \sqrt{f(z)}})\xi^{(1)}(z)=\frac{1+\xi^{(0)}(z)^2}{f(z)} .
\eea
The in-falling boundary conditions at the horizon implies  $\xi^{(0)}(z_H)=i,\, \xi^{(1)}(z_H)=0$. We  now can calculate the decay rate $\Gamma$ using
\bea
\Gamma=-\frac{\mathcal{A}^{(0)}}{\mathcal{A}^{(1)}}=-\frac{\textrm{Im} \xi^{(0)}(z)}{\textrm{Im} \xi^{(1)}(z)}\Big |_{z \to 0} .
\eea
\subsection* {case $m\neq0$}
A decay rate $\Gamma$ for  AdS$_{d+1}$  with $\Phi=M_0 z +M z^2$ is given by, 
\bea
\Gamma (T)=&\frac{2 \pi T}{ \int_{0}^{1} dt (1-t)^{(1/d-1)} {F(t)}/{t}},  \\
 &\textrm{ with } 
F(t)=-\sinh [ \frac{4m}{d}\tan^{-1}\!\!\sqrt{t}+\frac{g M_0 \beta_{0}(t)}{T}+ \frac{g M \beta_{1}(t)}{T^2 }], 
\label{gammaT}
\eea
where $\beta_{0}(t)=\frac{1}{2\pi}B(t;\frac{1}{2},\frac{1}{d}), \quad \beta_{1}(t)=\frac{d}{8 \pi^{2}}B(t;\frac{1}{2},\frac{2}{d})$. Notice that in this paper we take $-1/2<m<0$. This result explains the appearance of the strange metallicity at the critical point and near by region.  At the criticality where the order parameters are zero, 
the appearance of the strange metalicity is equivalent to the presence of the black hole horizon, which in turn is equivalent to the presence of the scrambling power of the chaotic fluctuation of the quantum critical point.  
However, the appearance of other phases is consequence of the symmetry breaking.  
Such competition of the strange metallicity, or chaos, and the order determines the shape of the 
phase diagram near the QCP. 
Notice that 
\be 
\Gamma\simeq\pi T/\gamma_{m,d}, \quad \hbox{ at }  T>>\sqrt{M}, M_{0},
\ee 
with 
$$\gamma_{m,d}^{-1}=\frac{2 m}{d} B(\frac{1}{2},\frac{1-2m}{d}) {}_3F_{2}(\frac{1}{2},\frac{1}{2}-\frac{2m}{d},1-\frac{2m}{d};\frac{3}{2},\frac{1}{2}+\frac{1-2m}{d};1).$$  For  $d=1, 2$, we have simple result $\gamma_d = \frac{\pi}{2} \tan(m \pi)$. We used  eq. (\ref{gammaT}) to calculate the phase boundaries.

\begin{figure}[ht!]
	\centering
	\subfigure[$M=-2$]	  
	{\includegraphics[width=6cm]{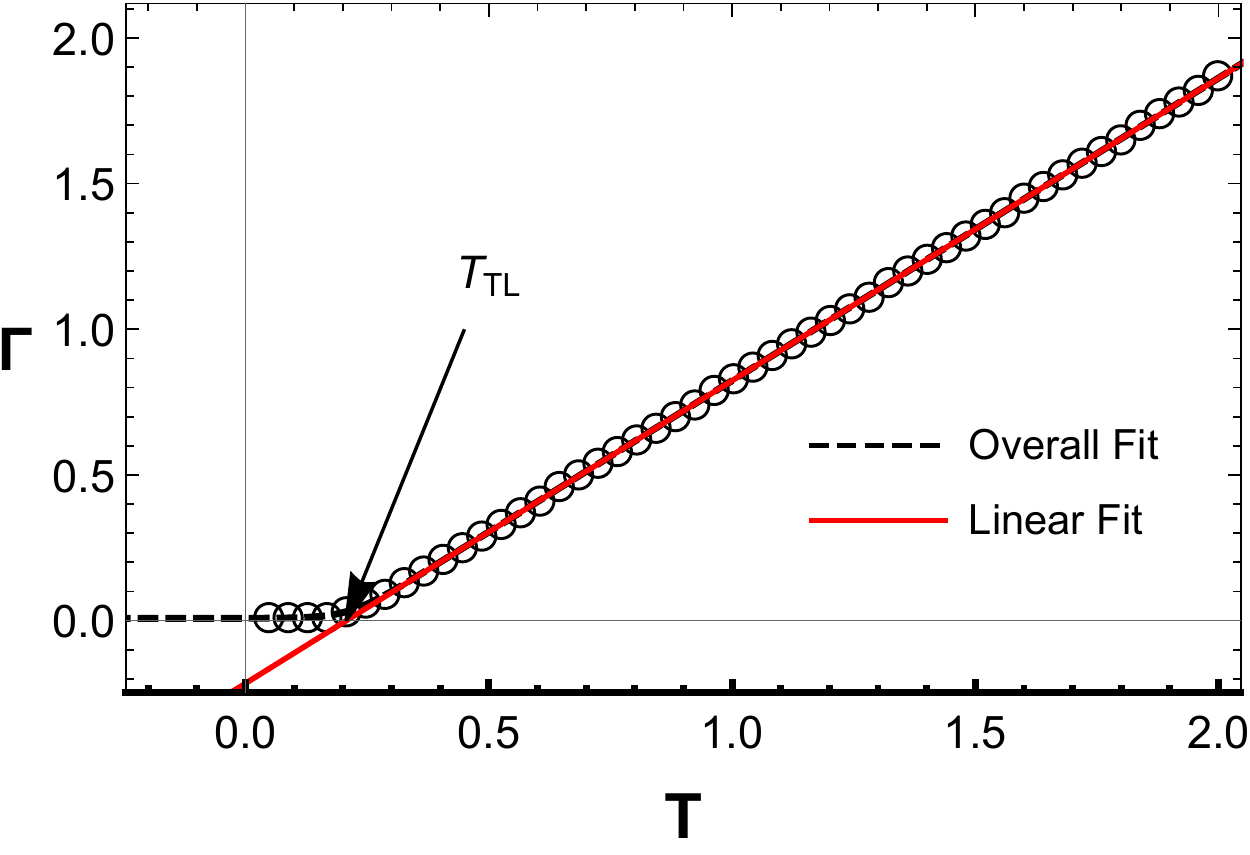}}
	\subfigure[$M=1$]
	{\includegraphics[width=6cm]{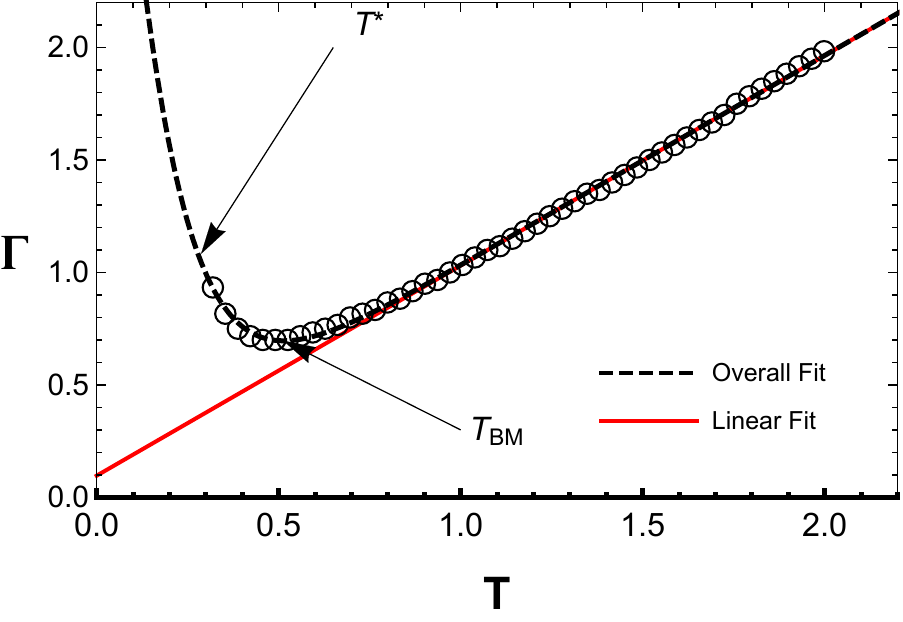}}
	\caption{$\Gamma(T)$ for $m=-0.4$:  For (a) $M=-2$ (b)$M=1$ respectively. Numerical $\Gamma$ is defined by a half width of spectral function at $w=0$}
	\label{check}
	\end{figure}

 One may want to compare    above analytic result with numerical calculation to check its validity. The result is Figure \ref{check}, showing that our formula agrees with numerical calculation precisely.   

\subsection*{  case $m=0$ }
For $m=0$, the spectral function $\mathcal{A}$ has non zero asymptotic value, i.e. $\lim_{\omega\to \infty} \Im G(\omega)=1$.  Therefore, the usual definition does not work. 
To overcome, we define the Drude function $\mathcal{A}_{D}$ by 
$\mathcal{A}_{D}=\mathcal{A} -1$. 
Then the relaxation time $\tau$ is defined by the small $\omega$ expansion 
$
	\mathcal{A}_{D}(w) \sim \mathcal{A}_{D}(0)(1- \tau w + \cdots), 
$
so that
$
\tau= \frac{1}{\mathcal{|A_D|}}\frac{\partial{\mathcal{A_D}}}{\partial{w}}|_{w=0}. 
$
Obviously negative-$\Gamma$ can be interpreted as a  measure of the gap. With this preparation,  we get the following formula for $m=0$ for $AdS_{d+1},d>1$: 
\bea
\Gamma=2 \pi T\frac{(e^{\frac{\alpha g M_0}{T}}-1)}{ \int_{0}^{1}(1-t)^{(1/d-1)}\frac{\sinh\left( g M_0 \beta_{0}(t)/T\right)}{t}dt},
\eea
where $\alpha= \frac{1}{2\pi}B(\frac{1}{2},\frac{1}{d}), \quad \beta_{0}(t)=\frac{1}{2\pi}B(t;\frac{1}{2},\frac{1}{d})$. Notice that  for the large T,   the width is also reduced to linear in $T$ : 
$\Gamma\simeq \frac{\pi T}{\gamma_d}$, with $\gamma_d={}_3F_2(\frac{1}{2},\frac{1}{2},1-\frac{1}{d};\frac{3}{2},\frac{1}{2}+\frac{1}{d};1)$.

Similarly, for the condensation, the decay rate with the scalar condensation is
\bea
\Gamma&=2 \pi T\frac{(e^{\frac{\alpha' g M}{T^2}}-1)}{ \int_{0}^{1}(1-t)^{(1/d-1)}\frac{\sinh\left( g M \beta'(t)/T^2\right)}{t}dt} 
\eea
where $\alpha'=\frac{d}{8 \pi^{2}}B(\frac{1}{2},\frac{2}{d}), \, \beta'(t)=\frac{d}{8 \pi^{2}}B(t;\frac{1}{2},\frac{2}{d}),$
 $\Gamma \simeq \frac{\pi T}{\gamma'_d} $ in the large T limit, with 
 $$\gamma'_d=\sqrt{\pi}{2^{1-\frac{2}{d}}}\frac{\Gamma(\frac{1}{2}+\frac{2}{d})}{{\Gamma(\frac{1}{2}+\frac{1}{d})}^2} \cdot {}_3F_2(\frac{1}{2},\frac{1}{2},1-\frac{2}{d};\frac{3}{2},\frac{1}{2}+\frac{2}{d};1) .$$  
 We tabulated $\gamma_d$ and $\gamma'_d$ in Table \ref{gammad} explicitly. 
\begin{table}[ht!]
	\centering
	\label{gammad1}
	\begin{tabular}{|c|c|c|c|c|} 
		\hline
		d & 1 & 2 & 3 & 4\\
		\hline\hline
		$\gamma_d$ & 1 & 1.1662 & $ {\sqrt{3} \pi}/{4}$ & $ {\pi}/{2}$\\
		\hline
		$\gamma'_d$ & $\frac{7}{5}$ & $ {\pi}/{2}$ & 1.7538 & 1.9468 \\
		\hline
	\end{tabular}
	\caption{$\gamma_d$ and $\gamma'_d$ for AdS$_{d+1},\quad d=1,2,3,4$.}
	\label{gammad}
\end{table}


  \section{Comment on the conformal factor $z^{\pm m}$}
The point we want to make below  is that $z^{\pm m}$ is not the  factor that counts probability of location but the conformal factor to embed  the  boundary theory  into the AdS bulk, which we should delete in probability interpretation of locality.  
To see why this is so,  we remind  the basic dictionary of the AdS/CFT for scalar case: if a scalar operator of dimension $\Delta$   couple    to the source $\phi_{0}(x)$, then the bulk  field dual to the operator, $\Phi(z,x)$,    is NOT given by the direct extension of the  $\phi_{0}$. Due to the conformal structure of AdS space\cite{Witten:1998qj}, we need to dress the source and response  by the conformal factor $z^{\Delta_{\mp}} $   such that two independent solutions of the bulk field $\Phi$   are given by 
    $\Phi(z,x)=\phi_{0}(x)z^{\Delta_{-}}(1+\cdots)$ and  $\Phi(z,x)= \vev{O}z^{\Delta_{+}}(1+\cdots)$, where $\Delta_{\pm}$ 
  are the solutions of $\Delta(\Delta-d)=m_{\Phi}^{2}$. 
The equation of motion for $\Phi$ is  of   second order  so that the general solution is given by the linear combination of the two. 
 Conversely,  in other to read off the `extended  configuration of   
 $\phi_{0}$'    from the solution of the bulk equation of motion,  we need to strip off the factor $z^{\Delta_{\pm}}$ from the first and second solutions. 
 Therefore, we define the undressed physical source/response  bulk field $ \Phi_{u,s}, \Phi_{u,r}$ by   dividing out the conformal factor 
 from the corresponding solutions of the bulk equation of motion. 

 Similarly for the spinor fields $\psi=(\psi_{+},\psi_{-})^{T}$,  
we can define the undressed source and response functions by $\psi_{\pm}= z^{\Delta_{\mp}}\psi_{u\pm}$ with $\Delta_{\pm}=3/2\pm m$.  It is   worthwhile to mention that the boundary Green functions are given by the ratio of these undressed wave functions for  both   bosons and  fermions. 
 Then we can show that  {   the  normalizable  undressed wave function  is localized at the boundary and  only   the zero mode has such property.}

\section{Localization of  zero modes at the AdS boundary}

Coming back to the AdS plus mirror space, 
the undressed wave function  for  the zero mode is 
 \be
 \psi_{0u +}\simeq e^{-M_{0}|z|}, \hbox{ for } -\infty <z<\infty . 
\ee  
This is precisely the Jackiw-Rebbi's normalizable soliton solution localized at the domain wall $z=0$ where the kink configuration is realized  by term $M_{0}{\rm sign} (z)$. 
 It also  means that our  zero mode 
can be considered as the edge state of a virtual topological insulator. 
However, we should not forget that our zero mode describe the bulk mode since it is free to move along the boundary of the AdS, which is the bulk of the physical world.  
In all these discussion, we introduced the mirror AdS to take direct similarity of JR solution. However   one should notice that all that  is used is to have the refection symmetry of the wave equation and a non-vanishing Dirichlet boundary condition of  the undressed wave function $\psi_{u\pm}$ at $z=0$. 
 
For  $\Phi=Mz^{2}$ and the equation of motion (\ref{FEM}) is invariant under $z\to -z$ without sign change of $M$.
The  ground state  given by 
\be
\psi_{ +}\simeq e^{- {M z^2}/{2}}, \ee
 which is the zero mode localized at the domain wall at $z=0$. 
 
\begin{figure}[ht!]
	\centering	  
	\subfigure[$g=1$]
	{\includegraphics[width=6cm]{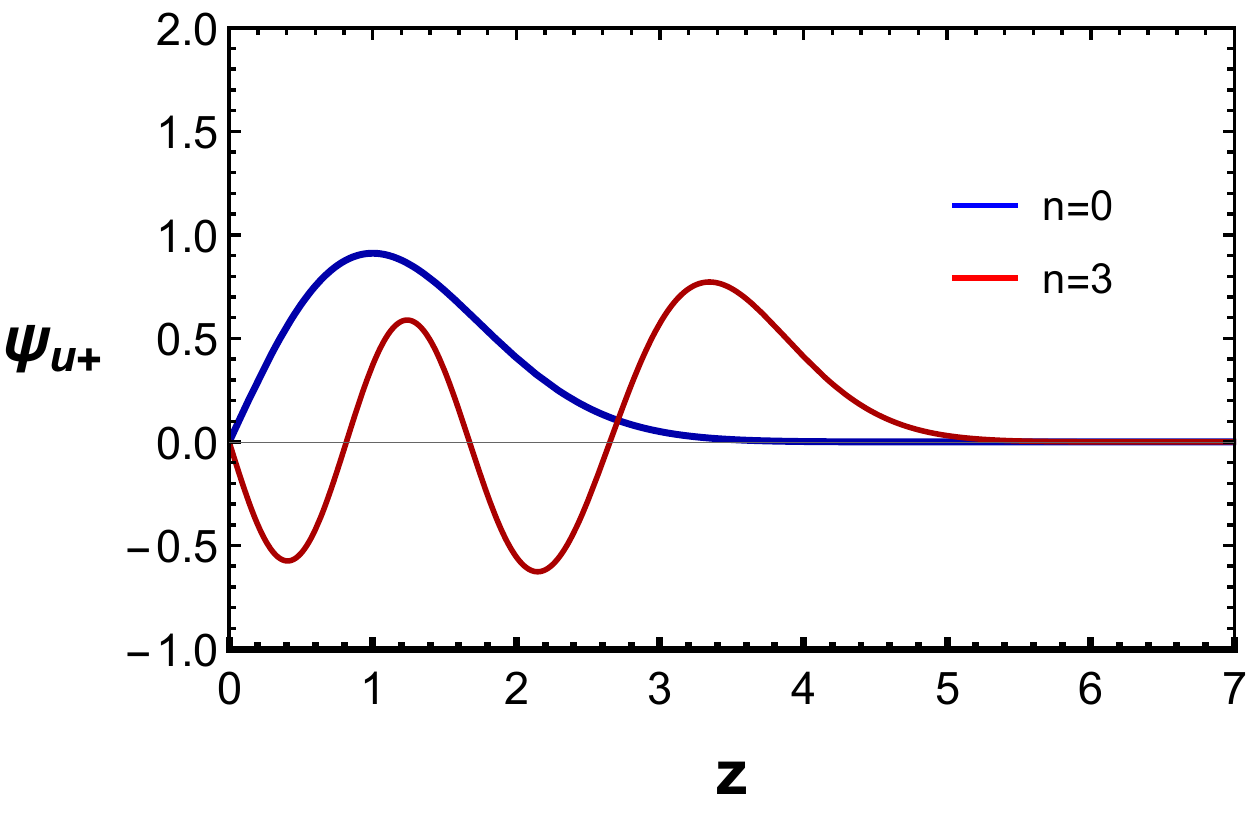}}
	\subfigure[$g=-1$]
	{\includegraphics[width=6cm]{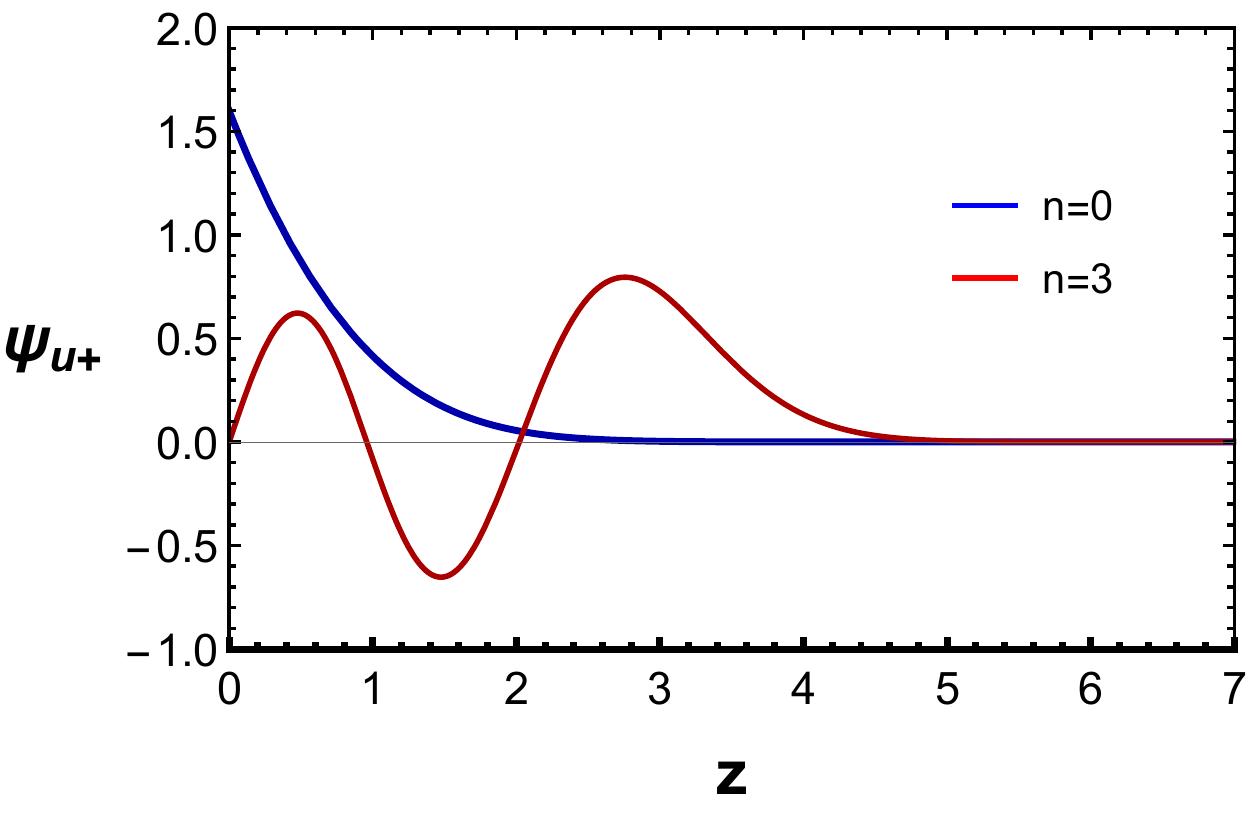}}
	\caption{wave function for $M=1, m=0$ for various modes. Only the ground state for $g=-1$ state is localized at the boundary of the AdS.}
	\label{wave}
\end{figure}

 One might think that  any state are localized  the boundary of the AdS. Below, we will show  only the ground state for  $g=-1$ is localized while  all other states are not.  To see this, notice that 
\bea
\psi^{g=1}_{u +}(z=0)&\sim 2^{\frac{1}{4}-\frac{m}{2}}\frac{\Gamma(\frac{1}{2}+m)}{\Gamma(-n)} =0, \hbox{ for any integer } n \geq 0  , \\
\psi^{g=-1}_{u +}(z=0)&\sim 2^{\frac{1}{4}-\frac{m}{2}}\frac{\Gamma(\frac{1}{2}+m)}{\Gamma(1-n)}>0, \hbox{ for  } n = 0 , 
\eea
which shows clearly  the uniqueness of  the zero mode  for $g=-1$. 
For a massive mode with $n\geq 1$, the wave functions are not localized for two reasons: first, the wave functions vanish at $z=0$ and secondly they oscillate  and penetrates into larger $z$ region. The larger is $n$, the deeper it penetrates. See the Figure 
\ref{wave}.
As far as $M$ or $M_{0}$, which is the gap between the ground state and the others,   is non-zero, there exists a fermion zero mode for $g=-1$ so we may say that the fermion zero mode is protected by the presence of the gap,  therefore we call the symmetry broken phase with $g=-1$ as  `topological liquid'. 
The   point we want to make is that the zero mode has a topological character,  and as such,  it has to do with 
a phenomena which looks like Fermi liquid   but with an unusual stability.  

\end{document}